\documentclass[preprint,review,12pt,3p,times]{elsarticle}

\usepackage{ifxetex}
\usepackage{ifpdf}
\ifxetex
  \usepackage[bookmarksnumbered,colorlinks,pagebackref]{hyperref}
\else
  \usepackage[CJKbookmarks=true,unicode,bookmarksnumbered,colorlinks,pagebackref]{hyperref}
\fi

\usepackage{amssymb}
\usepackage{amsmath}
\usepackage{amsthm}
\usepackage{bm}
\usepackage{color,xcolor}
\usepackage{epstopdf}
\usepackage{subfig}
\usepackage[ruled]{algorithm2e}
\usepackage{algorithm2e,setspace}
\usepackage{upgreek}
\usepackage{booktabs}
\usepackage{array}
\usepackage{makecell}
\usepackage{lineno}
\biboptions{sort&compress}

\newcommand{\refeq}[1]{(\ref{#1})}  % 引用公式形式：(#1)

\newcommand{\tabincell}[2]{
\begin{tabular}{@{}#1@{}}#2\end{tabular}
}

\journal{}

\begin{document}

\begin{frontmatter}



\title{High-accuracy three-dimensional surface detection in smoothed particle hydrodynamics for free-surface flows}

\author[USTC,IAPCM,GSCAEP]{Wen-Bin Liu}
\author[IAPCM]{Dong-Jun Ma}
\author[IAPCM]{Jian-Zhen Qian}
\author[IAPCM]{Ming-Yu Zhang}
\author[IAPCM]{An-Min He}
\author[USTC]{Nan-Sheng Liu}
\author[USTC,IAPCM,CAPT]{Pei Wang\corref{cor1}}

\address[USTC]{Department of Modern Mechanics, University of Science and Technology of China, Hefei, Anhui, 230027, China}
\address[IAPCM]{Institute of Applied Physics and Computational Mathematics, Beijing 100094, China}
\address[GSCAEP]{Graduate School of China Academy of Engineering Physics, Beijing 100088, China}
\address[CAPT]{Center for Applied Physics and Technology, Peking University, Beijing 100871, China}

\cortext[cor1]{Author to whom correspondence should be addressed: wangpei@iapcm.ac.cn}

\begin{abstract}
In this study, we investigate high-accuracy three-dimensional surface detection in smoothed particle hydrodynamics for free-surface flows. A new geometrical method is first developed to enhance the accuracy of free-surface particle detection in complex flows. This method detects free-surface particles via continuous global scanning inside the sphere of a particle through a cone region whose vertex corresponds to the particle position. The particle is identified as a free-surface particle if there exists a cone region with no neighboring particles. Next, an efficient semi-geometrical method is proposed based on the geometrical method to reduce the computational cost. It consists of finding particles near the free surface via position divergence and then detecting these particles using the geometrical method to identify free-surface particles. The accuracy and robustness of the proposed method are demonstrated by performing tests on several model problems.
\end{abstract}



\begin{keyword}
Free-surface particle detection \sep Geometrical method \sep Semi-geometrical method \sep Smoothed particle hydrodynamics




\end{keyword}

\end{frontmatter}


\section{Introduction}\label{sec:intro}

Smoothed particle hydrodynamics (SPH) is a representative Lagrangian particle method. In the SPH method, the materials are discretized into a series of particles with independent information. Because of its Lagrangian and meshless nature, the SPH method has the advantage of simulating free-surface and multi-phase flows with large deformations, hence, it is widely applied to scientific and engineering problems \cite{GingoldMonaghan1977, Monaghan2012, ZhangSun2017, YeLiu2019, ZhangLiu2018, VyasCummins2021}.

In many aspects of the SPH method, such as the implementation of boundary conditions \cite{ColagrossiAntuono2009, ColagrossiAntuono2011}, particle shifting technology (PST) \cite{SunColagrossi2017, HuangLong2019}, and continuum surface force (CSF) model \cite{Zhang2010, SunShen2019}, it is important to know exactly which particles belong to the free surface. Generally, two methods can be applied to detect free-surface particles: the kernel and geometrical methods \cite{Dilts2000,Morris2000, RandlesLibersky1996, SunShen2019, LinLiu2019, HaqueDilts2007, MarroneColagrossi2010, HuangZhang2018, WangMeng2019, SandimPaiva2020, ZhengSun2021}. In the kernel method, the sum of the kernel function or its gradients over the neighboring particles is used to detect free-surface particles \cite{Morris2000, RandlesLibersky1996}. Although the kernel method is easy to implement, its accuracy depends on the particle distribution. In the geometrical method, Dilts \cite{Dilts2000} first developed a two-dimensional exposure method. If the circle of a particle is not completely covered by the circles of neighboring particles, the particle is identified as a free-surface particle. Sun et al. \cite{SunShen2019} developed an improved two-dimensional geometrical method by performing continuous global scanning inside the circle of a particle through a sector region. The vertex of the sector corresponds to the particle position. If there exists a sector region with no neighboring particles, the particle is identified as a free-surface particle. Haque et al. \cite{HaqueDilts2007} extended the two-dimensional exposure method to be a three-dimensional version. If the sphere of a particle is not completely covered by the spheres of neighboring particles, the particle is identified as a free-surface particle. Marrone et al. \cite{MarroneColagrossi2010} proposed a semi-geometrical method for free-surface particle detection. The particles near the free surface are first found by the minimum eigenvalue of a renormalization matrix. Then, the free-surface particles are accurately detected by a scan region along the normal direction of the particle. Based on the method proposed by Marrone et al. \cite{MarroneColagrossi2010}, Wang et al. \cite{WangMeng2019} proposed an optimized semi-geometrical method for free-surface particle detection. This method makes use of the position divergence to find free-surface candidate particles. This considerably reduces the computational time for three-dimensional problems. However, owing to the estimation of the normal direction of the particle, both methods \cite{MarroneColagrossi2010, WangMeng2019} are sensitive to particle distribution.

To improve the accuracy of the free-surface particle detection in three-dimensional complex flows, we develop a new geometrical method for the three-dimensional SPH method, extending the two-dimensional geometrical method of Sun et al. \cite{SunShen2019}. Our method detects free-surface particles by performing continuous global scanning within the sphere of a particle through a cone region. The vertex of the cone corresponds to the particle position. If there exists a cone region with no neighboring particles, the particle is identified as a free-surface particle, otherwise, it is identified as an inner particle, i.e., there is at least one neighboring particle in any cone region. Next, based on the proposed geometrical method, a semi-geometrical method is developed to reduce computational costs. Finally, the proposed method is applied to several tests, and the results of the free-surface particle detection are compared with those of the previous methods \cite{HaqueDilts2007, MarroneColagrossi2010, WangMeng2019}.

This paper is organized as follows. In Section~\ref{sec:sph-eqn}, the SPH method for weakly-compressible flows is introduced. Then, in Section~\ref{sec:3d-sur}, we present a new method for free-surface particle detection. In Section~\ref{sec:results}, the proposed method is validated through four tests. Finally, the conclusions are drawn in Section~\ref{sec:conclusion}.

%===============================================================================
\section{SPH method}\label{sec:sph-eqn}

%-------------------------------------------------------------------------------
\subsection{Governing equations}

The Navier$-$Stokes equations for weakly-compressible fluids in Lagrangian description are given as follows:
\begin{equation}\label{eq:governing-equations}
  \left\{
  \begin{array}{l}
    \displaystyle\frac{D \rho}{Dt}
    =
    -\rho \nabla \cdot \boldsymbol{u}, \medskip \\
    \displaystyle\frac{D \boldsymbol{u}}{Dt}
    =
    -\displaystyle\frac{{1}}{{\rho}} \nabla p + \text{div} \mathbf{V} + \boldsymbol{g} + \frac{1}{\rho} \boldsymbol{f}^s,
  \end{array}
  \right.
\end{equation}
where $ \rho $ and $ p $ denote density and pressure, respectively. The $ \boldsymbol{u} $, $ \boldsymbol{g} $, and $ \boldsymbol{f}^s $ represent velocity vector, gravitational acceleration, and surface tension force, respectively. $ \mathbf{V} $ denotes the viscous stress tensor.

A linear state equation is used to close Eqs.~\refeq{eq:governing-equations} are as follows:
\begin{equation}
  p = c_{0}^{2} \left( \rho - \rho_0 \right),
\end{equation}
where $ \rho_0 $ and $ c_{0} $ denote the density and speed of sound under rest condition, respectively.

For weakly-compressible fluids \cite{MarroneColagrossi2015}, the speed of sound needs to satisfy the following conditions:
\begin{equation}
  c_0 \geq 10 \max \left( \sqrt{\frac{p_{\max}}{\rho_0}}, U_{\max} \right),
\end{equation}
where $ p_{\max} $ and $ U_{\max} $ denote the expected maximum pressure and velocity, respectively.

%-------------------------------------------------------------------------------
\subsection{{\rm SPH} equations}

The $\delta$-SPH method \cite{AntuonoColagrossi2010} is used to discretize the governing equations, in which a diffusive term for stabilizing the pressure field is added to the continuity equation. The discrete governing equations with diffusive terms are given as follows:
\begin{equation} \label{eq:delta-sph-model}
  \left\{
  \begin{array}{l}
    \displaystyle \frac{D \rho_i}{Dt}
    =
    -\rho_i \sum\limits_{j = 1}^N{\left( \boldsymbol{u}_j - \boldsymbol{u}_i \right) \cdot \nabla_i W_{ij} V_j} + \delta h_i c_0 D_i, \medskip \\
	\displaystyle \frac{D \boldsymbol{u}_i}{Dt}
    =
    \displaystyle -\frac{1}{\rho_i} \sum\limits_{j = 1}^N{\left( p_i + p_j \right) \nabla_i W_{ij} V_j} + \displaystyle \frac{1}{\rho_i} 2 \left( d + 2 \right) \mu \sum\limits_{j = 1}^N{\pi_{ij} \nabla_i W_{ij} V_j} + \boldsymbol{g} + \frac{1}{\rho_i} \boldsymbol{f}_{i}^{s},
  \end{array}
  \right.
\end{equation}
where $ N $, $ d $, and $ V $ denote the number of neighboring particles, spatial dimensions, and volume of a particle, respectively. The parameter $ \delta $ is set to 0.1 \cite{AntuonoColagrossi2012}.

The term $ D_i $ in the continuity equation is given as follows:
\begin{equation}
  \left\{
  \begin{array}{l}
    \displaystyle D_i
    =
    2 \sum\limits_{j = 1}^N{\psi_{ij} \frac{\left( \boldsymbol{x}_j - \boldsymbol{x}_i \right) \cdot \nabla_i W_{ij}}{\left| \boldsymbol{x}_j - \boldsymbol{x}_i \right|^2}}, \medskip \\
	\displaystyle \psi_{ij}
    =
    \left( \rho_j - \rho_i \right) - \frac{1}{2}\left( \left< \nabla \rho \right>_{i}^{L} + \left< \nabla \rho \right>_{j}^{L} \right) \cdot \left( \boldsymbol{x}_j - \boldsymbol{x}_i \right),
  \end{array}
  \right.
\end{equation}
where $ \boldsymbol{x}_i $ denotes the particle position vector. $ \left< \nabla \rho \right> ^L $ is expressed as:
\begin{equation}
  \left\{
  \begin{array}{l}
	\left< \nabla \rho \right> _{i} ^{L}
    =
    \displaystyle \sum\limits_{j = 1}^N{\left( \rho_j - \rho_i \right) \mathbf{L}_i \nabla_i W_{ij} V_j}, \medskip \\
	\mathbf{L}_i
    =
    \displaystyle \left[ \sum\limits_{j = 1}^N{\left( \boldsymbol{x}_j - \boldsymbol{x}_i \right) \otimes \nabla_i W_{ij} V_j} \right] ^{-1}.
  \end{array}
  \right.
\end{equation}

The term $ \pi_{ij} $ in the momentum equation is given as follows:
\begin{equation}
  \pi_{ij}
  =
  \frac{\left( \boldsymbol{u}_j - \boldsymbol{u}_i \right) \cdot \left( \boldsymbol{x}_j - \boldsymbol{x}_i \right)}{\left| \boldsymbol{x}_j - \boldsymbol{x}_i \right| ^2}.
\end{equation}

The Wendland C2 kernel function is used, which is written as follows:
\begin{equation}
  W \left( \left| \boldsymbol{x}_j - \boldsymbol{x}_i \right|, h \right)
  =
  \left\{
  \begin{array}{ll}
    \displaystyle \alpha_d \left( 1 + 2q \right) \left( 1 - \frac{q}{2} \right)^4,
    &
    0 \leq q \leq 2 \medskip \\
    \displaystyle 0,
    &
    q > 2
  \end{array}
  \right.
\end{equation}
where $ q = \left| \boldsymbol{x}_j - \boldsymbol{x}_i \right|/h $ and $ h $ denotes the smoothing length. The parameters $ \alpha_d $ is given as follows:
\begin{equation}
  \alpha_d
  =
  \left\{
  \begin{array}{ll}
    \displaystyle \frac{3}{4h},
    &
    d = 1 \medskip \\
	\displaystyle \frac{7}{4 \pi h^2},
    &
    d = 2 \medskip \\
	\displaystyle \frac{21}{16 \pi h^3},
    &
    d = 3
  \end{array}
  \right.
\end{equation}

The PST developed by Wang et al. \cite{WangMeng2019} is used to improve the spatial configuration of the particles and is given as follows:
\begin{equation}
  \boldsymbol{x}_{i}^{*} = \boldsymbol{x}_i + \delta \boldsymbol{x}_i,
\end{equation}
\begin{equation}
  \delta \boldsymbol{x}_i
  =
  \left\{
  \begin{array}{lll}
    0
    &
    &
    {\rm if} \;  i \in F \medskip\\
	-k_{\rm CFL} \cdot {\rm Ma} \cdot 2h^2 \cdot \displaystyle \sum\limits_{j=1}^N{\frac{m_j}{\rho_j} R \left( \frac{W_{ij}}{W \left( \Delta x_i \right)} \right)^n \nabla_i W_{ij}}
    &
    j \in \left[ \left| \boldsymbol{x}_{ij} \right| < l_i \right]
    &
    {\rm if} \; i \in V \medskip\\
	-k_{\rm CFL} \cdot {\rm Ma} \cdot 2h^2 \cdot \displaystyle \sum\limits_{j=1}^N{\frac{m_j}{\rho_j} \left[ 1 + R \left( \frac{W_{ij}}{W \left( \Delta x_i \right)} \right)^n \right] \nabla_i W_{ij}}
    &
    j \in \left[ \left| \boldsymbol{x}_{ij} \right| < kh \right]
    &
    {\rm if} \; i \in I
  \end{array}
  \right.
\end{equation}
where $ \delta \boldsymbol{x}_i $ and $ \boldsymbol{x}_{i}^{*} $ denote the particle shifting vector and new particle position vector, respectively. $ F $, $ V $, and $ I $ denote the free-surface region, free-surface vicinity region, and inner region, respectively. $ n $, $ R $, and $ k_{\rm CFL} $ are set to 4, 0.2, and 0.25, respectively.

%===============================================================================
\section{High-accuracy method for three-dimensional free-surface particle detection}\label{sec:3d-sur}

This section demonstrates the high-accuracy method for three-dimensional free-surface particle detection in the SPH method. First, a high-accuracy geometrical method for three-dimensional free-surface particle detection is developed. Continuous global scanning within the sphere of a particle is performed through a cone region, rather than along the normal direction alone, which is difficult in ensuring accurate computation of complex flows. The particle is identified as a free-surface particle if there exists a cone region with no neighboring particles. Then, based on the proposed geometrical method, a semi-geometrical method is developed to reduce computational costs.

The symbols employed in this section are defined as follows. Let $ S_i $ denote the sphere with radius $ R $ centered at particle $ i $, and $ C_j $ denote the cone of rotation angle $ \theta $ and rotation axis $ l_{ij} $ with particle $ i $ as the vertex, where $ l_{ij} $ is the line passing through particle $ i $ and its neighboring particle $ j $. Moreover, $ C_{ij} $ is used to denote the circle obtained by intersecting the sphere $ S_i $ with cone $ C_j $. See Fig.~\ref{fig:schematic-diagram} for details.

\subsection{Geometrical method for free-surface particle detection} \label{subsec:geometrical-method}

\begin{figure}[htbp]
\centering
\includegraphics[width=16cm]{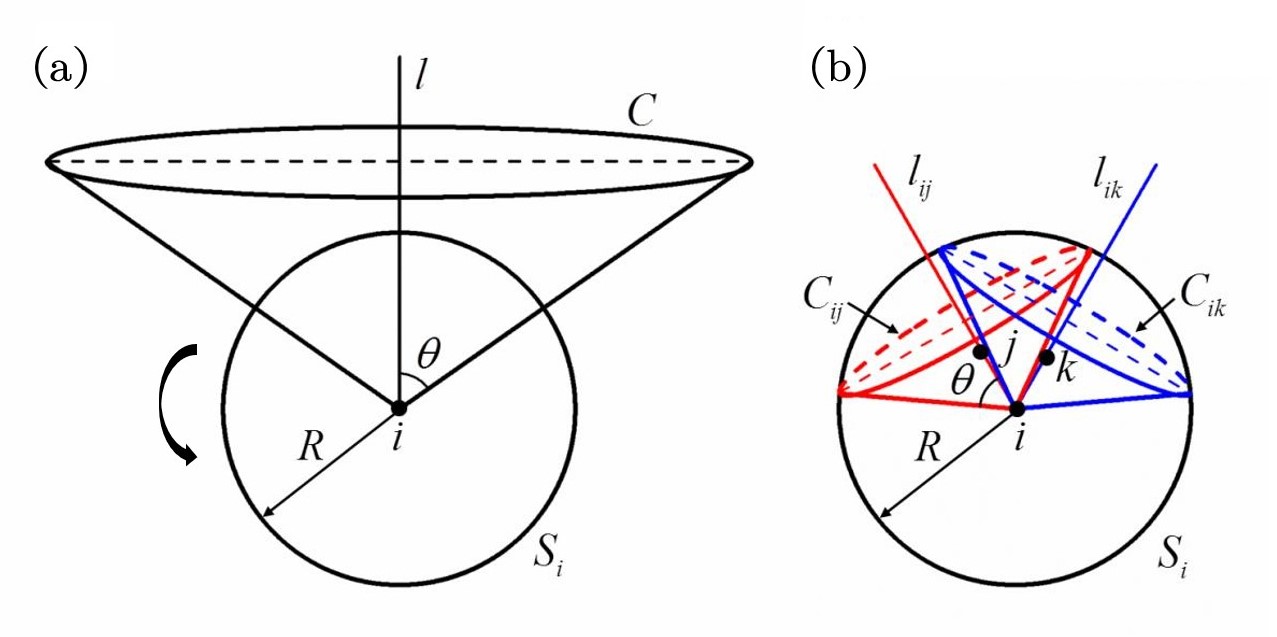}
\caption{Black circle denotes candidate free-surface sphere $ S_i $ with radius $ R $ centered at particle $ i $. (a) Black cone denotes scanning cone of rotation angle $ \theta $ and rotation axis $ l $ with particle $ i $ as the vertex. (b) Red cone denotes cone $ C_j $ of rotation angle $ \theta $ and rotation axis $ l_{ij} $ with particle $ i $ as the vertex, where $ l_{ij} $ is the line passing through particle $ i $ and neighboring particle $ j $. Blue cone denotes cone $ C_k $ of rotation angle $ \theta $ and rotation axis $ l_{ik} $ with particle $ i $ as the vertex, where $ l_{ik} $ is the line passing through particle $ i $ and neighboring particle $ k $. Red $ C_{ij} $ denotes the circle given by the intersection of sphere $ S_i $ and cone $ C_j $. Blue $ C_{ik} $ denotes the circle given by the intersection of sphere $ S_i $ and cone $ C_k $.}
\label{fig:schematic-diagram}
\end{figure}

The key idea of the geometrical method is to perform continuous global scanning inside the sphere $ S_i $ through a cone region $ C $. The vertex of the cone $ C $ corresponds to the position of particle $ i $, and the rotation angle of cone $ C $ is $ \theta $. If there exists a cone region with no neighboring particles, the particle $ i $ is identified as a free-surface particle; otherwise, it is identified as an inner particle, i.e., there is at least one neighboring particle in any cone region, as shown in Fig.~\ref{fig:schematic-diagram}a. Because the number of such cones is infinite during the continuous global scanning, the aforementioned idea cannot be directly implemented in the program. However, this idea can be equivalently expressed as follows: the sphere $ S_i $ intersects the cone $ C_j $ of all neighboring particles and the circles $ C_{ij} $ are drawn on the surface of sphere $ S_i $. If the surface of sphere $ S_i $ is not completely covered by circles $ C_{ij} $, i.e., some parts of the circle $ C_{ij} $ are exposed, the particle $ i $ is identified as a free-surface particle; otherwise, it is identified as an inner particle, as shown in Fig.~\ref{fig:schematic-diagram}b. The logical proof of the aforementioned equivalence transformation is given as follows: when the surface of sphere $ S_i $ is not completely covered by the circles $ C_{ij} $, for any point $ m $ on the surface of sphere $ S_i $ not completely covered by circles $ C_{ij} $, the angle of ray $ l_{im} $ and all neighboring rays $ l_{ij} $ is greater than $ \theta $, and there must be no neighboring particles in the cone $ C_m $ with rotation angle $ \theta $ and rotation axis $ l_{im} $. Therefore, the particle $ i $ is identified as a free-surface particle. In contrast, when the surface of sphere $ S_i $ is completely covered by the circles $ C_{ij} $, for any point $ m $ on the surface of sphere $ S_i $, it must be within a circle $ C_{ij} $, and the angle of ray $ l_{im} $ and $ l_{ij} $ is less than $ \theta $, then there is at least one neighboring particle in the cone $ C_m $ with the rotation angle $ \theta $ and the rotation axis $ l_{im} $. Therefore, the particle $ i $ is identified as an inner particle. Algorithm~\ref{algo:geometrical-method} gives the main operations of the present geometrical method. The remainder of this section focuses on the following functions:\\
\begin{itemize}
  \item Intersection of sphere and cone
  \item Intersection of two circles
  \item Checking coverage of circle
\end{itemize}

\begin{algorithm}
\setstretch{1.1}
\SetAlgoLined
\DontPrintSemicolon
\For{$ i $ {\rm over all particles}}
{
  \For{$ j $ {\rm over neighboring particles of particle} $ i $}
  {
    $ C_{ij} $ = \emph{Intersection of sphere and cone} ($ S_i, C_j $)\;
    \For{$ k $ {\rm over neighboring particles of particle} $ i $}
    {
      \If{{\rm angle} $ \varphi_{ijk} $ {\rm of rays} $ l_{ij} $ {\rm and} $ l_{ik} $ {\rm is greater than} $ 0^\circ $ {\rm and less than} $ 2 \theta $}
      {
        Add the particle $ k $ to the set $ \mathcal{L} $
      }
    }
    Sort the set $ \mathcal{L} $ from smallest to largest $ \varphi_{ijk} $\;
    \For{$ k $ {\rm over the particles of set} $ \mathcal{L} $}
    {
      $ C_{ik} $ = \emph{Intersection of sphere and cone} ($ S_i, C_k $)\;
      \emph{Intersection of two circles} ($ C_{ij}, C_{ik} $)\;
      \emph{Checking coverage of circle}\;
      \If{$ C_{ij} $ {\rm is completely covered by the circles} $ C_{ik} $}
      {
        $ \mathbf{continue} $ with the next $ j $
      }
    }

    \uIf{$ C_{ij} $ {\rm is not completely covered by the circles} $ C_{ik} $}
    {
      Particle $ i $ is identified as the free-surface particle\;
      $ \mathbf{continue} $ with the next $ i $
    }
    \Else
    {
      $ \mathbf{continue} $ with the next $ j $
    }
  }
  \uIf{{\rm for every} $ j $, {\rm the} $ C_{ij} $ {\rm is completely covered by the circles} $ C_{ik} $}
  {
    Particle $ i $ is identified as the inner particle
  }
  \Else
  {
    Particle $ i $ is identified as the free-surface particle
  }
}
\caption{Geometrical method for free-surface particle detection}
\label{algo:geometrical-method}
\end{algorithm}

\subsubsection{Intersection of sphere and cone}

Let $ \boldsymbol{x}_i $ = ($ x_i, y_i, z_i $) denote the position of particle $ i $, which is also the center of sphere $ S_i $. Likewise, $ \boldsymbol{x}_j $ = ($ x_j, y_j, z_j $) denotes the position of the neighboring particle $ j $. Hence, the distance $ d_{ij} $ between particles $ i $ and $ j $ is given as follows:
\begin{equation}
  d_{ij}
  =
  \sqrt{\left( x_j - x_i \right)^2 + \left( y_j - y_i \right)^2 + \left( z_j - z_i \right)^2}.
\end{equation}

\begin{figure}[htbp]
\centering
\includegraphics[width=10cm]{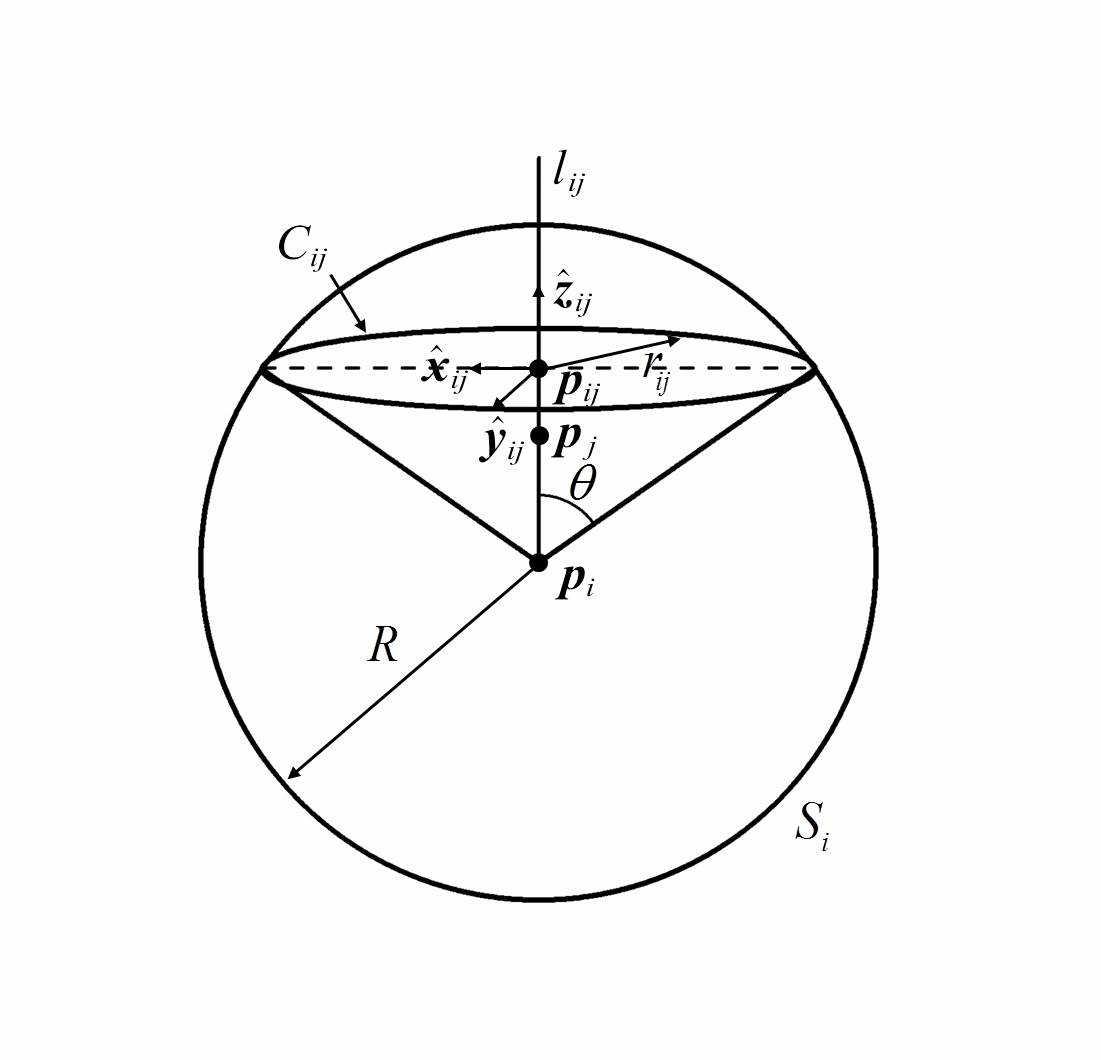}
\caption{Center $ \boldsymbol{x}_{ij} $, radius $ r_{ij} $ and local coordinate system for the circle $ C_{ij} $.}
\label{fig:zcztic-diagram}
\end{figure}

As shown in Fig.~\ref{fig:zcztic-diagram}, if $ d_{ij} \le R $, the center $ \boldsymbol{x}_{ij} $ = ($ x_{ij}, y_{ij}, z_{ij} $) and radius $ r_{ij} $  of circle $ C_{ij} $ are given as follows:
\begin{equation}
  \boldsymbol{x}_{ij}
  =
  \boldsymbol{x}_i
  +
  \frac{R \cos \theta}{d_{ij}} \left( \boldsymbol{x}_j - \boldsymbol{x}_i \right),
\end{equation}
\begin{equation}
  r_{ij}
  =
  R \sin \theta.
\end{equation}

To easily calculate the angle interval intersected by two circles, a local coordinate system is established, in which $ \boldsymbol{x}_{ij} $ is the origin, and the unit axis vectors $ \boldsymbol{\hat{x}}_{ij} $, $ \boldsymbol{\hat{y}}_{ij} $, and $ \boldsymbol{\hat{z}}_{ij} $  are given as follows:
\begin{equation}
  \boldsymbol{\hat{z}}_{ij}
  =
  \frac{1}{d_{ij}} \left( \boldsymbol{x}_j - \boldsymbol{x}_i \right),
\end{equation}
\begin{equation}
  \boldsymbol{\hat{x}}_{ij}
  =
  \left\{
  \begin{array}{ll}
	\displaystyle \frac{\boldsymbol{\hat{x}} \times \boldsymbol{\hat{z}}_{ij}} {\lVert \boldsymbol{\hat{x}} \times \boldsymbol{\hat{z}}_{ij} \rVert},
    &
    \; \text{if} \;\; \boldsymbol{\hat{y}}\times \boldsymbol{\hat{z}}_{ij} = 0 \medskip \\
	\displaystyle \frac{\boldsymbol{\hat{y}} \times \boldsymbol{\hat{z}}_{ij}} {\lVert \boldsymbol{\hat{y}} \times \boldsymbol{\hat{z}}_{ij} \rVert},
    &
    \; \text{if} \;\; \boldsymbol{\hat{y}}\times \boldsymbol{\hat{z}}_{ij} \ne 0
  \end{array}
  \right.
\end{equation}
\begin{equation}
  \boldsymbol{\hat{y}}_{ij}
  =
  \displaystyle \frac{\boldsymbol{\hat{z}}_{ij} \times \boldsymbol{\hat{x}}_{ij}} {\lVert \boldsymbol{\hat{z}}_{ij} \times \boldsymbol{\hat{x}}_{ij} \rVert},
\end{equation}
where $ \boldsymbol{\hat{x}} $ and $ \boldsymbol{\hat{y}} $ are the unit vectors of the original coordinate system, respectively. The unit vector $ \boldsymbol{\hat{z}}_{ij} = \left( z_{ij}^{\left( 1 \right)}, z_{ij}^{\left( 2 \right)}, z_{ij}^{\left( 3 \right)} \right) $  is also the outward normal vector of the circle $ C_{ij} $.

\subsubsection{Intersection of two circles}

Define $ \varphi_{ijk} = \cos^{-1} \left( \boldsymbol{\hat{z}}_{ij} \cdot \boldsymbol{\hat{z}}_{ik} \right) $ as the angle of the unit vectors $ \boldsymbol{\hat{z}}_{ij} $ and $ \boldsymbol{\hat{z}}_{ik} $. Here, four possible scenarios can be derived:

(1) $ \varphi_{ijk} = 0^\circ $, circles $ C_{ij} $ and $ C_{ik} $ coincide.

(2) $ 0^\circ < \varphi_{ijk} < 2 \theta $ , circles $ C_{ij} $ and $ C_{ik} $ intersect at two points.

(3) $ \varphi_{ijk} = 2 \theta $ , circles $ C_{ij} $ and $ C_{ik} $ intersect at one point.

(4) $ \varphi_{ijk} > 2 \theta $ , circles $ C_{ij} $ and $ C_{ik} $ do not intersect.

In cases 1 and 3, circles $ C_{ij} $ and $ C_{ik} $ are considered to have no intersection.

Before calculating the angle intervals $ A_{ijk} $ of the intersection of circles $ C_{ij} $ and $ C_{ik} $, the intersection particles are sorted based on the $ \varphi_{ijk} $ from smallest to largest. The smaller the angle of $ \varphi_{ijk} $, the greater the angle intervals $ A_{ijk} $, in this case, the possibility of fully covering circle $ C_{ij} $ increases rapidly. Once the circle $ C_{ij} $ is determined as covered, no more angle intervals $ A_{ijk} $ need to be computed.

\begin{figure}[htbp]
\centering
\includegraphics[width=10cm]{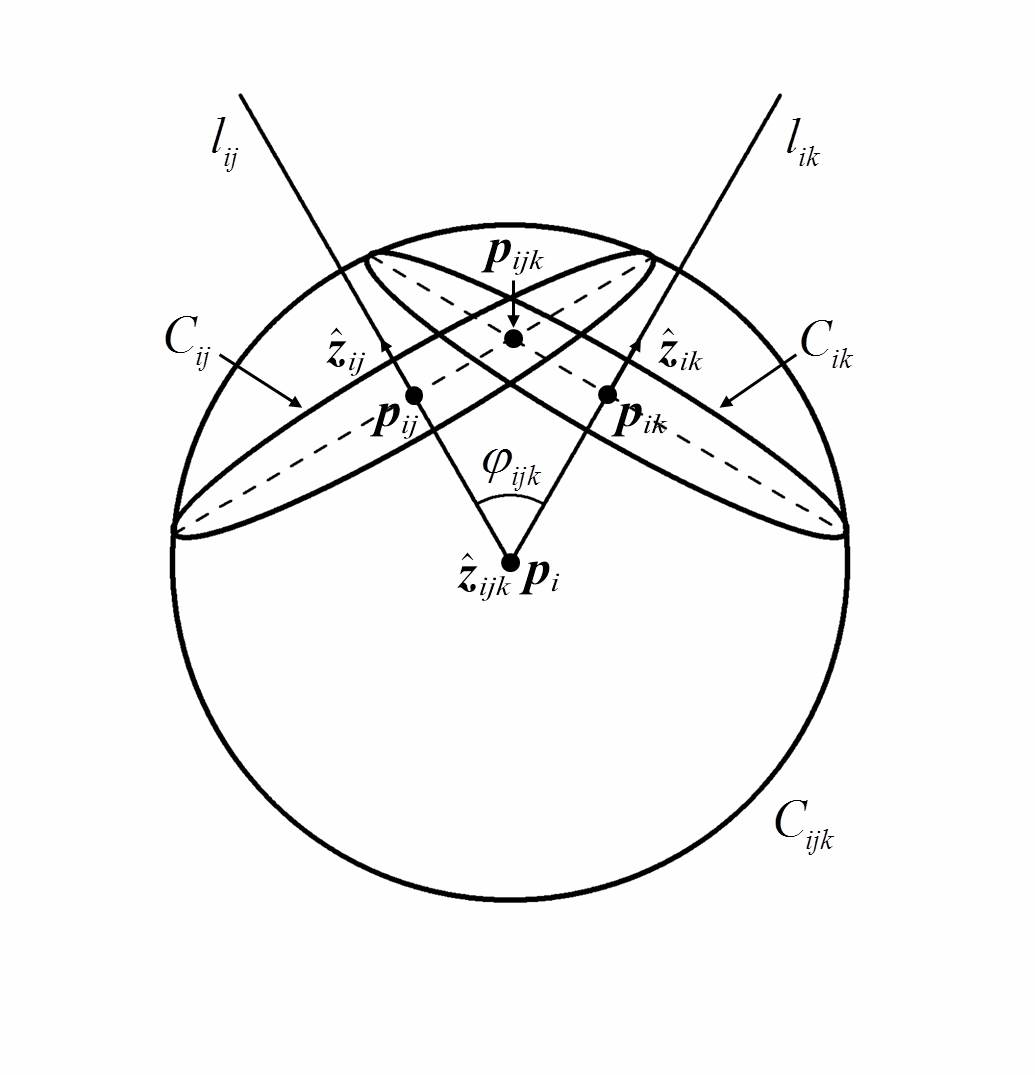}
\caption{Intersection of two circles. The point $ \boldsymbol{x}_{ijk} $ is the intersection of the planes of circles $ C_{ij} $, $ C_{ik} $ and $ C_{ijk} $.}
\label{fig:intersection-circles}
\end{figure}

For case 2 where circles $ C_{ij} $ and $ C_{ik} $ intersect at two points, we consider the plane containing the points $ \boldsymbol{x}_i $, $ \boldsymbol{x}_{ij} $ and $ \boldsymbol{x}_{ik} $. Its intersection with sphere $ S_i $ is the circle $ C_{ijk} $, as shown in Fig.~\ref{fig:intersection-circles}. The unit normal vector  $ \boldsymbol{\hat{z}}_{ijk} = \left( z_{ijk}^{\left( 1 \right)}, z_{ijk}^{\left( 2 \right)}, z_{ijk}^{\left( 3 \right)} \right) $ of circle $ C_{ijk} $ is:
\begin{equation}
  \boldsymbol{\hat{z}}_{ijk}
  =
  \displaystyle \frac{\boldsymbol{\hat{z}}_{ij} \times \boldsymbol{\hat{z}}_{ik}} {\lVert \boldsymbol{\hat{z}}_{ij} \times \boldsymbol{\hat{z}}_{ik} \rVert}.
\end{equation}

The equations describing the planes of the three circles $ C_{ij} $, $ C_{ik} $, and $ C_{ijk} $ in the point-direction are:
\begin{equation}
  \left\{
  \begin{array}{l}
	\boldsymbol{\hat{z}}_{ij}\cdot \left( \boldsymbol{x} - \boldsymbol{x}_{ij} \right) = 0 , \smallskip \\
	\boldsymbol{\hat{z}}_{ik}\cdot \left( \boldsymbol{x} - \boldsymbol{x}_{ik} \right) = 0 , \smallskip \\
	\boldsymbol{\hat{z}}_{ijk}\cdot \left( \boldsymbol{x} - \boldsymbol{x}_{i} \right) = 0,
  \end{array}
  \right.
\end{equation}
where $ \boldsymbol{x} = \left( x, y, z \right) $. The above equations can be simplified as:
\begin{equation}
  \left\{
  \begin{array}{l}
	z_{ij}^{\left( 1 \right)} x + z_{ij}^{\left( 2 \right)} y + z_{ij}^{\left( 3 \right)} z
    =
    z_{ij}^{\left( 1 \right)} x_{ij} + z_{ij}^{\left( 2 \right)} y_{ij} + z_{ij}^{\left( 3 \right)} z_{ij}, \medskip \\
	z_{ik}^{\left( 1 \right)} x + z_{ik}^{\left( 2 \right)} y + z_{ik}^{\left( 3 \right)} z
    =
    z_{ik}^{\left( 1 \right)} x_{ik} + z_{ik}^{\left( 2 \right)} y_{ik} + z_{ik}^{\left( 3 \right)} z_{ik}, \medskip \\
	z_{ijk}^{\left( 1 \right)} x + z_{ijk}^{\left( 2 \right)} y + z_{ijk}^{\left( 3 \right)} z
    =
    z_{ijk}^{\left( 1 \right)} x_{i} + z_{ijk}^{\left( 2 \right)} y_{i} + z_{ijk}^{\left( 3 \right)} z_{i}.
  \end{array}
  \right.
\end{equation}

The point $ \boldsymbol{x}_{ijk} = \left( x_{ijk}, y_{ijk}, z_{ijk} \right) $ ,  which is the intersection of the planes of circles $ C_{ij} $, $ C_{ik} $, and $ C_{ijk} $, can be obtained using:
\begin{equation}
  \left[
  \begin{array}{l}
	x_{ijk} \smallskip \\
	y_{ijk} \smallskip \\
	z_{ijk}
  \end{array}
  \right]
  =
  \left[
  \begin{array}{lll}
	z_{ij}^{\left( 1 \right)} & z_{ij}^{\left( 2 \right)} & z_{ij}^{\left( 3 \right)} \medskip \\
	z_{ik}^{\left( 1 \right)} & z_{ik}^{\left( 2 \right)} & z_{ik}^{\left( 3 \right)} \medskip \\
	z_{ijk}^{\left( 1 \right)}& z_{ijk}^{\left( 2 \right)}& z_{ijk}^{\left( 3 \right)}
  \end{array}
  \right]^{-1}
  \left[
  \begin{array}{l}
	z_{ij}^{\left( 1 \right)} x_{ij} + z_{ij}^{\left( 2 \right)} y_{ij} + z_{ij}^{\left( 3 \right)} z_{ij}  \medskip \\
	z_{ik}^{\left( 1 \right)} x_{ik} + z_{ik}^{\left( 2 \right)} y_{ik} + z_{ik}^{\left( 3 \right)} z_{ik}  \medskip \\
	z_{ijk}^{\left( 1 \right)} x_i   + z_{ijk}^{\left( 2 \right)} y_i   + z_{ijk}^{\left( 3 \right)} z_i
  \end{array}
  \right].
\end{equation}

The local coordinates of point $ \boldsymbol{x}_{ijk}^{*} = \left( x_{ijk}^{*}, y_{ijk}^{*}, z_{ijk}^{*} \right) $ in the plane of circle $ C_{ij} $ are:
\begin{equation}
  \left\{
  \begin{array}{l}
	x_{ijk}^{*}
    =
    \left( x_{ijk} - x_{ij}, y_{ijk} - y_{ij}, z_{ijk} - z_{ij} \right) \cdot \boldsymbol{\hat{x}}_{ij}, \medskip \\
    y_{ijk}^{*}
    =
    \left( x_{ijk} - x_{ij}, y_{ijk} - y_{ij}, z_{ijk} - z_{ij} \right) \cdot \boldsymbol{\hat{y}}_{ij}, \medskip \\
	z_{ijk}^{*}
    =
    0.
  \end{array}
  \right.
\end{equation}

\begin{figure}[htbp]
\centering
\includegraphics[width=10cm]{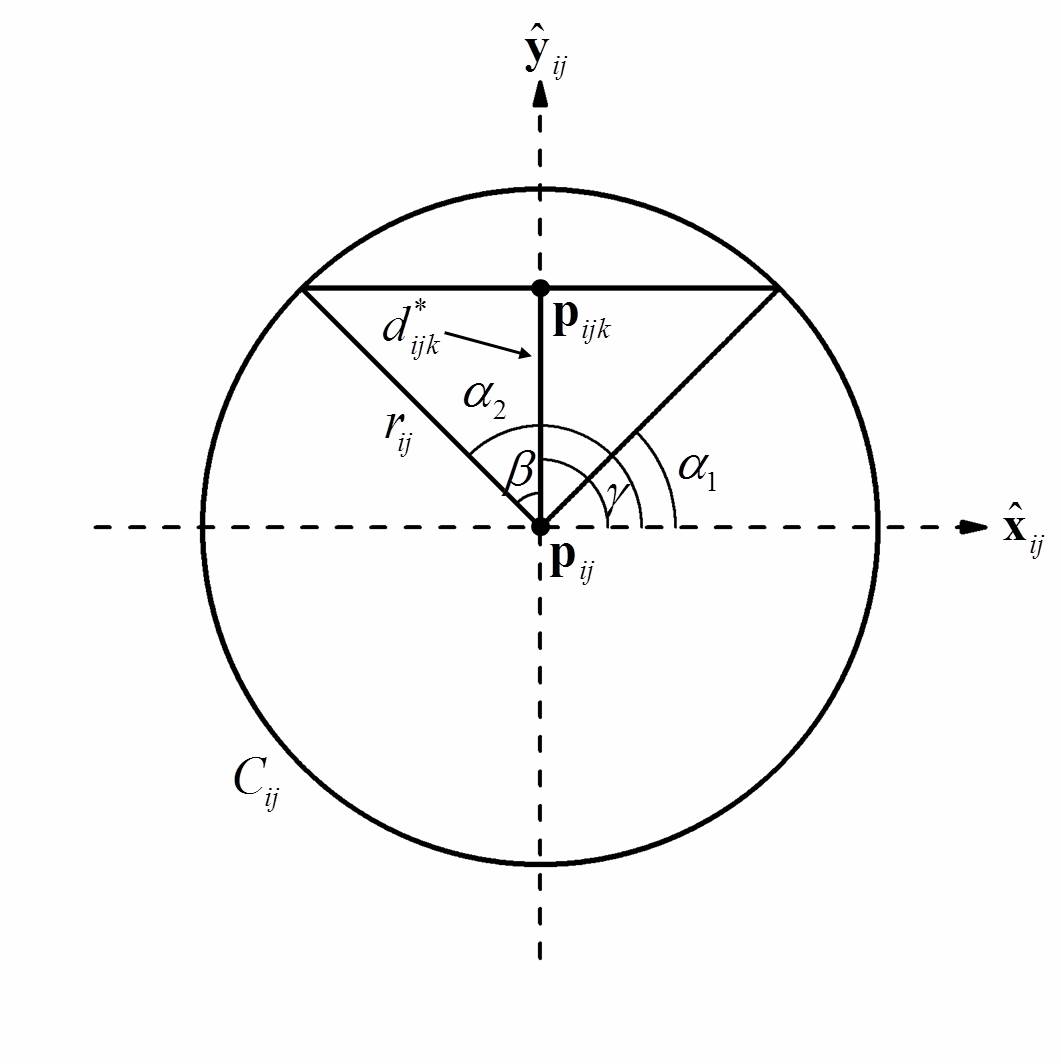}
\caption{Finding the angle interval $ A_{ijk} $ of the intersection of circles $ C_{ij} $ and $ C_{ik} $.}
\label{fig:angle-interval}
\end{figure}

Consider the plane of circle $ C_{ij} $, as shown in Fig.~\ref{fig:angle-interval}. The angles $ \beta $ and $ \gamma $ are:
\begin{equation}
  \beta
  =
  \cos^{-1} \left( \frac{d_{ijk}^{*}}{r_{ij}} \right),
\end{equation}
\begin{equation}
  \gamma
  =
  \left\{
  \begin{array}{ll}
	\cos^{-1} \left( \displaystyle \frac{x_{ijk}^{*}}{d_{ijk}^{*}} \right),
    &
    \; \text{if} \;\; y_{ijk}^{*} \geq 0 \medskip \\
	2 \pi - \cos^{-1} \left( \displaystyle \frac{x_{ijk}^{*}}{d_{ijk}^{*}} \right),
    &
    \; \text{if} \;\; y_{ijk}^{*} < 0
  \end{array}
  \right.
\end{equation}
where $ d_{ijk}^{*} = \sqrt{x_{ijk}^{*} \cdot x_{ijk}^{*} + y_{ijk}^{*} \cdot y_{ijk}^{*}} $. The angle interval $ A_{ijk} $ of the intersection of circles $ C_{ij} $ and $ C_{ik} $ is:
\begin{equation}
  A_{ijk}
  =
  \left[ \alpha_1, \alpha_2 \right]
  =
  \left[ \gamma - \beta, \gamma + \beta \right].
\end{equation}

\subsubsection{Checking coverage of circle}

\begin{figure}[htbp]
\centering
\includegraphics[width=12cm]{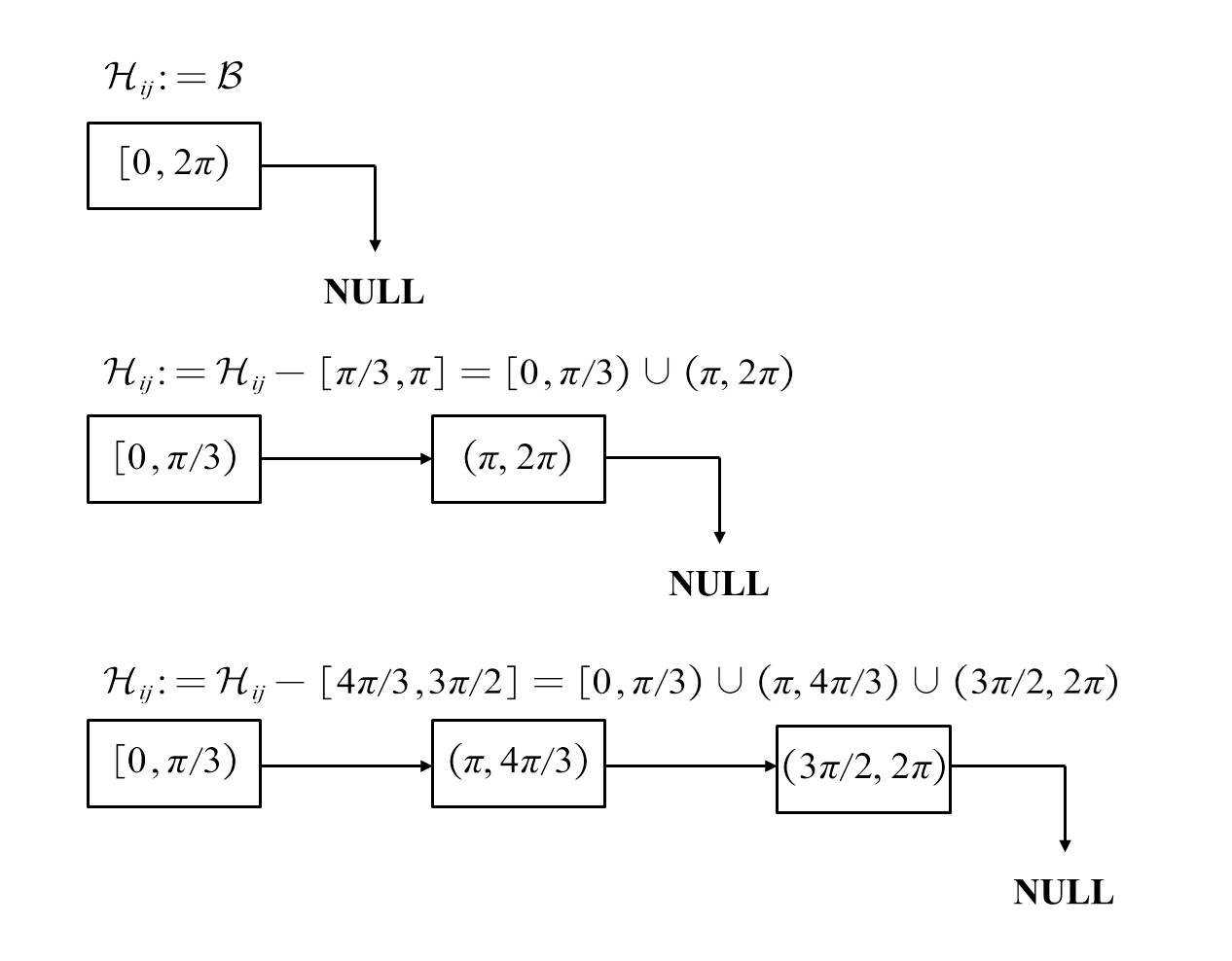}
\caption{Example of the dynamic list update procedure.}
\label{fig:sketch-linkedlist}
\end{figure}

After obtaining the angle intervals $ A_{ijk} $ of the intersection between circles $ C_{ij} $ and $ C_{ik} $, the checking coverage of a circle is accomplished using the method developed by Haque et al. \cite{HaqueDilts2007}. This method employs a dynamic list that represents the parts of the circle $ C_{ij} $ that have not been intersected by the neighboring circles $ C_{ik} $. When the dynamic list is empty, it can be concluded that the circle $ C_{ij} $ is entirely covered by the neighboring circles $ C_{ik} $ and the calculation of the angle intervals $ A_{ijk} $ can be terminated. Figure~\ref{fig:sketch-linkedlist} shows an example of the dynamic list update procedure, in which the dynamic list $ \mathcal{H}_{ij} $ is initialized to the entire interval $ \mathcal{B} = \left[ 0,2\pi \right) $ and represents the complement of the angle interval $ \tilde{A}_{ijk} $ in the current list $ \mathcal{H}_{ij} $ in the following calculations, where it is expressed as:
\begin{equation}
  \mathcal{H}_{ij} = \mathcal{H}_{ij} - \tilde{A}_{ijk},
\end{equation}
where $ \tilde{A}_{ijk} $ is expressed as follows:
\begin{equation}
  \tilde{A}_{ijk}
  =
  \left\{
  \begin{array}{lll}
   	\left[ 0, \alpha_2 \right] \cup \left[ \alpha_1 + 2\pi, 2 \pi \right)
    &
    \text{if}
    &
    \, \alpha_1 < 0 \medskip \\
 	\left[ 0, \alpha_2 - 2\pi \right] \cup \left[ \alpha_1, 2 \pi \right)
    &
    \text{if}
    &
    \, \alpha_2 \geq 2 \pi  \medskip \\
	A_{ijk}
    &
    \text{else}
    &

  \end{array}
  \right.
\end{equation}

When the linked list is empty, the circle $ C_{ij} $ is completely covered by the angle intervals $ A_{ijk} $:
\begin{equation}
  \mathcal{H}_{ij} = \oslash \Longleftrightarrow C_{ij} \; \text{is covered}
\end{equation}

Because the intersection particles are sorted from smallest to largest $ \varphi_{ijk} $, the corresponding angle intervals $ A_{ijk} $ are obtained from largest to smallest. In fact, for most interior circles $ C_{ij} $, only a few angle intervals $ A_{ijk} $ are required to determine whether the circle $ C_{ij} $ is covered.

\subsubsection{Analysis of the geometrical method}\label{subsubsec:analysis}

The performance of the proposed geometrical method is determined by the radius of the sphere and the rotation angle of the cone. If the radius of the sphere is too large, the small hole inside the fluid will not be detected, and a huge detection time will be consumed. If it is too small, several inner particles will be misidentified as free-surface particles. In this study, the radius of the sphere is set to $ 2.5 \Delta x $. If the rotation angle of the cone is too large, the detection accuracy of the concave surface will decrease, and a huge detection time will be consumed. If it is too small, several inner particles will be misidentified as a free-surface particles. In this study, the rotation angle of the cone is set to $ 45^{\circ} $.

\subsection{Semi-geometrical method for free-surface particle detection}

The geometrical method proposed in Section~\ref{subsec:geometrical-method} can accomplish free-surface particle detection of the three-dimensional model using only the particle coordinates. However, if the proposed geometrical method is directly employed in a numerical simulation, it will result in high computation costs because it is necessary for each particle to perform continuous global scanning through a cone region at each time step. As a result, a semi-geometrical method is proposed to reduce computational costs based on the proposed geometrical method. The semi-geometrical method for detecting free-surface particles comprise the following two steps.

\begin{figure}[htbp]
\centering
\includegraphics[width=10cm]{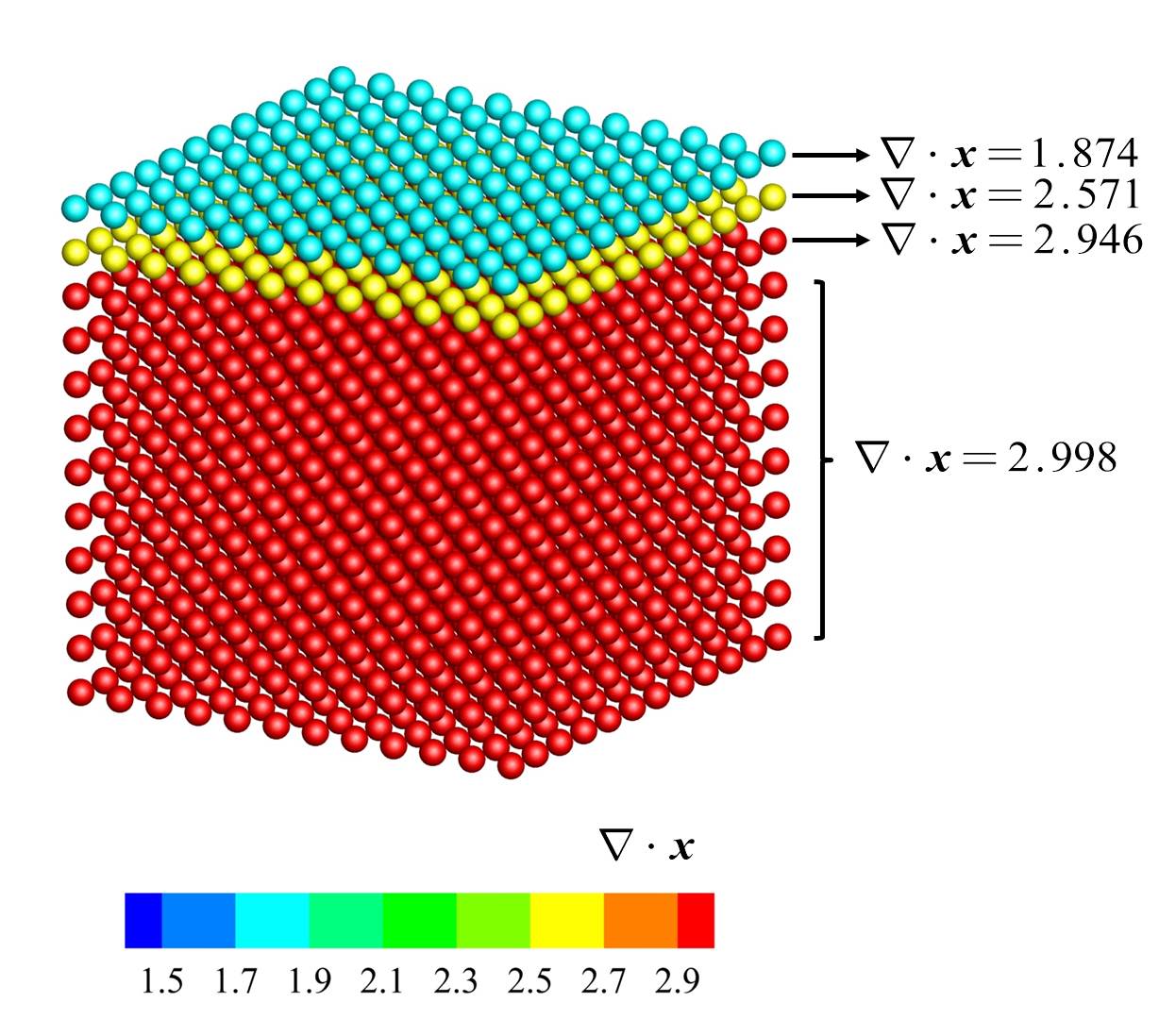}
\caption{The position divergences of particles in three dimensions.}
\label{fig:position-divergences}
\end{figure}

In the first step, the particles near the free surface are found by position divergence \cite{WangMeng2019}, which considerably reduces the number of particles to be accurately detected in the second step. The position divergence $ \lambda_i $ is calculated as follows:
\begin{equation}
    \lambda_i
    =
    \nabla \cdot \boldsymbol{x}_i
    =
    \sum_{j = 1}^N{\frac{m_j}{\rho_j} \left( \boldsymbol{x}_j - \boldsymbol{x}_i \right) \cdot \nabla_i W_{ij}}.
\end{equation}
Figure~\ref{fig:position-divergences} demonstrates the position divergence of particles in three dimensions. The position divergence of the free-surface particles is 1.874 when the particles are uniformly distributed. As a result, particles with a position divergence of less than 1.874 are directly identified as free-surface particles. In the proposed semi-geometrical method, we selected 1.8 as the threshold for detecting free-surface particles.
The position divergence of the particles in the two layers near the free surface ranges from 1.874 to 2.571. Therefore, particles with a position divergence exceeding 2.571 are directly identified as inner particles. We selected 2.8 as the threshold for detecting inner particles in the semi-geometrical method. Finally, for particles with a position divergence greater than 1.8 but less than 2.8, an accurate detection of the free-surface particle is executed in the second step. Note that the thresholds vary with respect to the ratio $ h / \Delta x $ and kernel function.

The second step involves accurate detection executed on free-surface candidate particles with a position divergence greater than 1.8 but less than 2.8. In contrast to the method used in the study \cite{MarroneColagrossi2010, WangMeng2019}, i.e., scanning the ``umbrella-shaped'' zone along the normal direction of the particle, continuous global scanning within the sphere $ S_i $ is accomplished via a cone region $ C $. If there exists a cone region with no neighboring particles, particle $ i $ is identified as a free-surface particle, otherwise, it is identified as an inner particle. Finally, the conditions for determining whether particle $ i $ is identified as a free-surface particle are given by:
\begin{equation}
  \left\{
  \begin{array}{lll}
	\lambda_i < 1.8
    &
    \Longrightarrow
    &
    i \in F \medskip \\
	1.8 \leq \lambda_i \leq 2.8 \; \& \& \; \exists \; C = \varnothing
    &
    \Longrightarrow
    &
    i \in F \medskip \\
	\text{otherwise}
    &
    \Longrightarrow
    &
    i \notin F
  \end{array}
  \right.
\end{equation}
where $ C $ denotes the scanning cone, and $ F $ denotes the free-surface zone. The performance of the proposed semi-geometrical method is tested in section~\ref{subsec:freesurface} and section~\ref{subsec:dambreaking}.

%===============================================================================
\section{Results and discussion}\label{sec:results}

Four tests are selected in this section to illustrate the present method for 3D free-surface particle detection. The advantages of the present method are analyzed in detail in the first two tests, i.e., surface detection of the free surface with periodic perturbations and dam-breaking, where the detection results and detection time are compared with those of the Haque method \cite{HaqueDilts2007}, the Marrone method \cite{MarroneColagrossi2010}, and the Wang method \cite{WangMeng2019}. Then, the present method is applied to the visualization tools and the particle shifting technology.

In the Haque method, the radius of the sphere is set to $ 1.25 \Delta x $, therefore, the spheres of neighboring particles within $ 2.5 \Delta x $ from the candidate particle intersect the sphere of the candidate particle, and the number of the neighboring particles is 81, which is the same as the present method. In the Marrone method, the renormalized Gaussian kernel function is used, and the smoothing length is set to $ 1.33 \Delta x $ \cite{MarroneColagrossi2010}. The QR decomposition method is used to solve the minimum eigenvalue of the renormalization matrix $ \mathbf{L}^{-1} $. The rotation angle of the ``umbrella-shaped'' zone in the second step is $ 45^{\circ} $. In the Wang method, the improved Gaussian kernel function is adopted, and the smoothing length is set to $ 1.5 \Delta x $ \cite{WangMeng2019}. The rotation angle of the ``umbrella-shaped'' zone in the second step is $ 45^{\circ} $. In the present method, the radius of the sphere is set to $ 2.5 \Delta x $, and the rotation angle of the cone is set to $ 45^{\circ} $.

\subsection{Surface detection of the free surface with periodic perturbations} \label{subsec:freesurface}

In this section, the present method is applied to the free surface with periodic perturbations and the detection results are compared with those of the Haque method \cite{HaqueDilts2007}, the Marrone method \cite{MarroneColagrossi2010}, and the Wang method \cite{WangMeng2019}. The distance between the free-surface particles is maintained as the initial inter-particle distance at varied angles of surface perturbation in addition to a uniform distribution of inner particles. The 3D detection methods are transformed into the corresponding 2D detection methods and applied to the 2D free surface with periodic perturbations to facilitate the analysis and presentation of the detection results. The Haque 3D method corresponds to the Dilts 2D method \cite{Dilts2000}, i.e., if the circle of a candidate particle is not completely covered by the circles of its neighboring particles, the candidate particle is identified as a free-surface particle. The Marrone method is applicable to both 2D and 3D cases, i.e., the particles near the free surface are initially found by the minimum eigenvalue of the renormalization matrix $ \mathbf{L}^{-1} $, then, the free-surface particles are accurately detected by a scan region along the normal direction of the candidate particle. The Wang method is applicable to both 2D and 3D cases, i.e., the particles near the free surface are initially found by the position divergence, then, the free-surface particles are accurately detected by a scan region along the normal direction of the candidate particle. The present 3D method corresponds to the Sun 2D method \cite{SunShen2019}, i.e., continuous global scanning within the circle of a candidate particle is performed through a sector region. If there exists a sector region with no neighboring particles, the candidate particle is identified as a free-surface particle.

\begin{figure}[htbp]
\centering
\includegraphics[width=16cm]{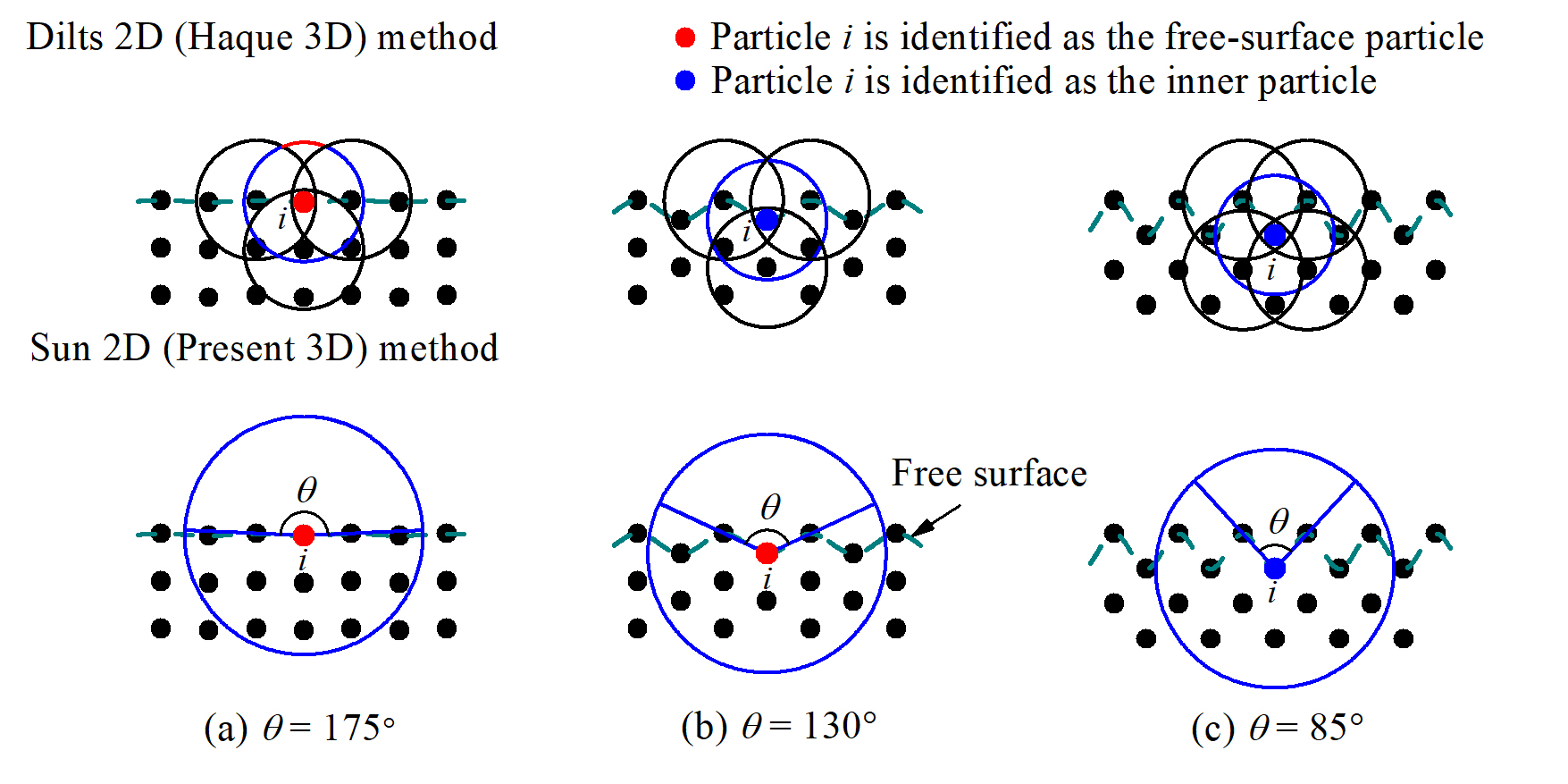}
\caption{Detection results of the Dilts 2D (Haque 3D) method and the Sun 2D (Present 3D) method applied to the 2D free surface with periodic perturbations, where the angles of the surface perturbations are $ 175^{\circ} $, $ 130^{\circ} $, and $ 85^{\circ} $.}
\label{fig:Surface-perturbations-1}
\end{figure}

\begin{figure}[htbp]
\centering
\includegraphics[width=16cm]{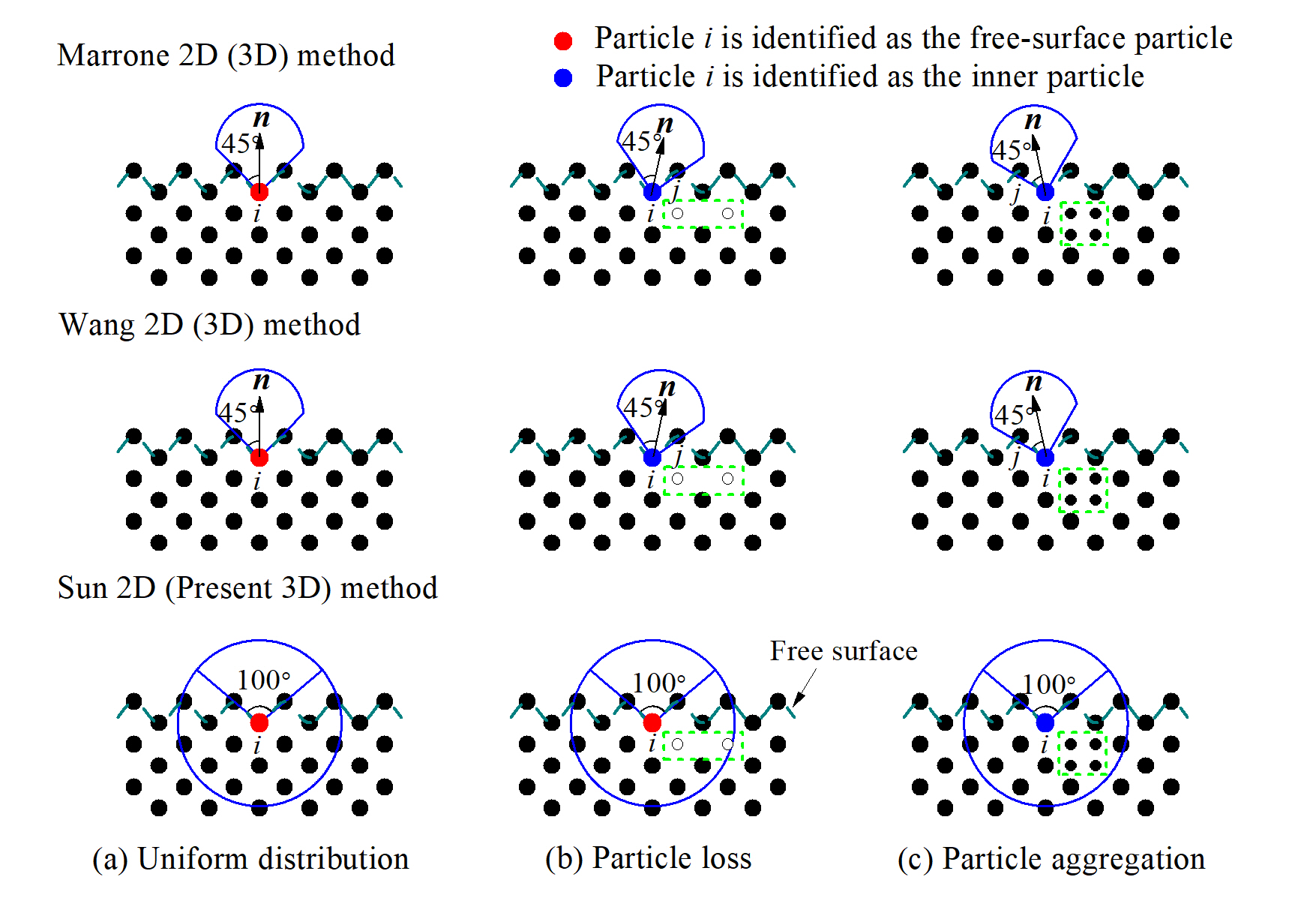}
\caption{Detection results of the Marrone 2D (3D) method, the Wang 2D (3D) method, and the Sun 2D (Present 3D) method applied to the 2D free surface with periodic perturbations for different inner particle distributions, where the angles of the surface perturbations are $ 100^{\circ} $.}
\label{fig:Surface-perturbations-2}
\end{figure}

Figure~\ref{fig:Surface-perturbations-1} shows the detection results of the Dilts method and the Sun method applied to the 2D free surface with periodic perturbations, where the angles of the surface perturbations are $ 175^{\circ} $, $ 130^{\circ} $, and $ 85^{\circ} $. In the Dilts method, a portion of the candidate circle is exposed, thus, the candidate particle $ i $ at the trough is identified as a free-surface particle when the angle of the surface perturbations is $ 175^{\circ} $, however, because the entire arc of the candidate circle is not exposed, the candidate particle $ i $ at the trough is identified as an inner particle when the angles of the surface perturbations are $ 130^{\circ} $ and $ 85^{\circ} $. In uniform particle distributions, the minimum angle $ \theta_{\min} $ that can be detected by the Dilts method is
\begin{equation}\label{eq:minimum-angle}
  \theta_{\min} = 2\arccos {(\frac{\Delta x}{2 R})},
\end{equation}
where $ R $ is the radius of the circle. When the radius of the circle is $ 1.25 \Delta x $, the minimum angle $ \theta_{\min} $ is $ 133^{\circ} $, where $ \theta_{\min} $ increases as the radius of the circle increases. In the Sun method, the angle of the scanning sector is set to $ 90^{\circ} $. When the angles of the surface perturbations are $ 175^{\circ} $ and $ 130^{\circ} $ which are greater than $ 90^{\circ} $, because continuous global scanning inside the circle of the candidate particle is performed through a sector region, there must exist a scanning sector region of particle $ i $ with no neighboring particles. Therefore, the candidate particle $ i $ at the trough is identified as a free-surface particle. When the angle of the surface perturbations is $ 85^{\circ} $ that is less than $ 90^{\circ} $, there is at least one neighboring particle in any scanning sector region of particle $ i $. Therefore, the candidate particle $ i $ at the trough is identified as an inner particle. To summarize, compared with the Dilts 2D method, the detection accuracy of the Sun 2D method is improved because of the enhanced capability to detect concave curves. Similarly, compared with the Haque 3D method, the detection accuracy of the present 3D method is improved because of the enhanced capability to detect concave surfaces.

Figure~\ref{fig:Surface-perturbations-2} depicts the detection results of the Marrone method, the Wang method, and the Sun method applied to the 2D free surface with periodic perturbations at various inner particle distributions, where the angles of the surface perturbations are $ 100^{\circ} $. In the Marrone method and the Wang method, because the scan region along the accurate normal direction does not contain neighboring particles, the candidate particle $ i $ is accurately identified as a free-surface particle when the inner particles are uniformly distributed. However, when maintaining a constant distribution of free-surface particles, an inaccurate estimation of the normal vector is obtained when the inner particles are lost or aggregated, and the scan region along the inaccurate normal direction contains a neighboring particle $ j $ such that the candidate particle $ i $ at the trough is misidentified as an inner particle. In the Sun method, the angle of the scanning sector is set to $ 90^{\circ} $. When the angle of the surface perturbation is $ 100^{\circ} $ that is greater than $ 90^{\circ} $, because continuous global scanning inside the circle of the candidate particle is performed through a sector region, there must exist a scanning sector region of particle $ i $ with no neighboring particles inside it. Therefore, the candidate particle $ i $ at the trough is identified as a free-surface particle at various inner particle distributions. To summarize, compared with the Marrone 2D method and the Wang 2D method, the detection accuracy of the Sun 2D method is improved because the detection accuracy is independent of the estimation of the normal vector of particles. Similarly, compared with the Marrone 3D method and the Wang 3D method, the detection accuracy of the present 3D method is improved because the detection accuracy is independent of the estimation of the normal vector of particles.

\subsection{Surface detection of the dam-breaking} \label{subsec:dambreaking}

\begin{figure}[htbp]
\centering
\includegraphics[width=16cm]{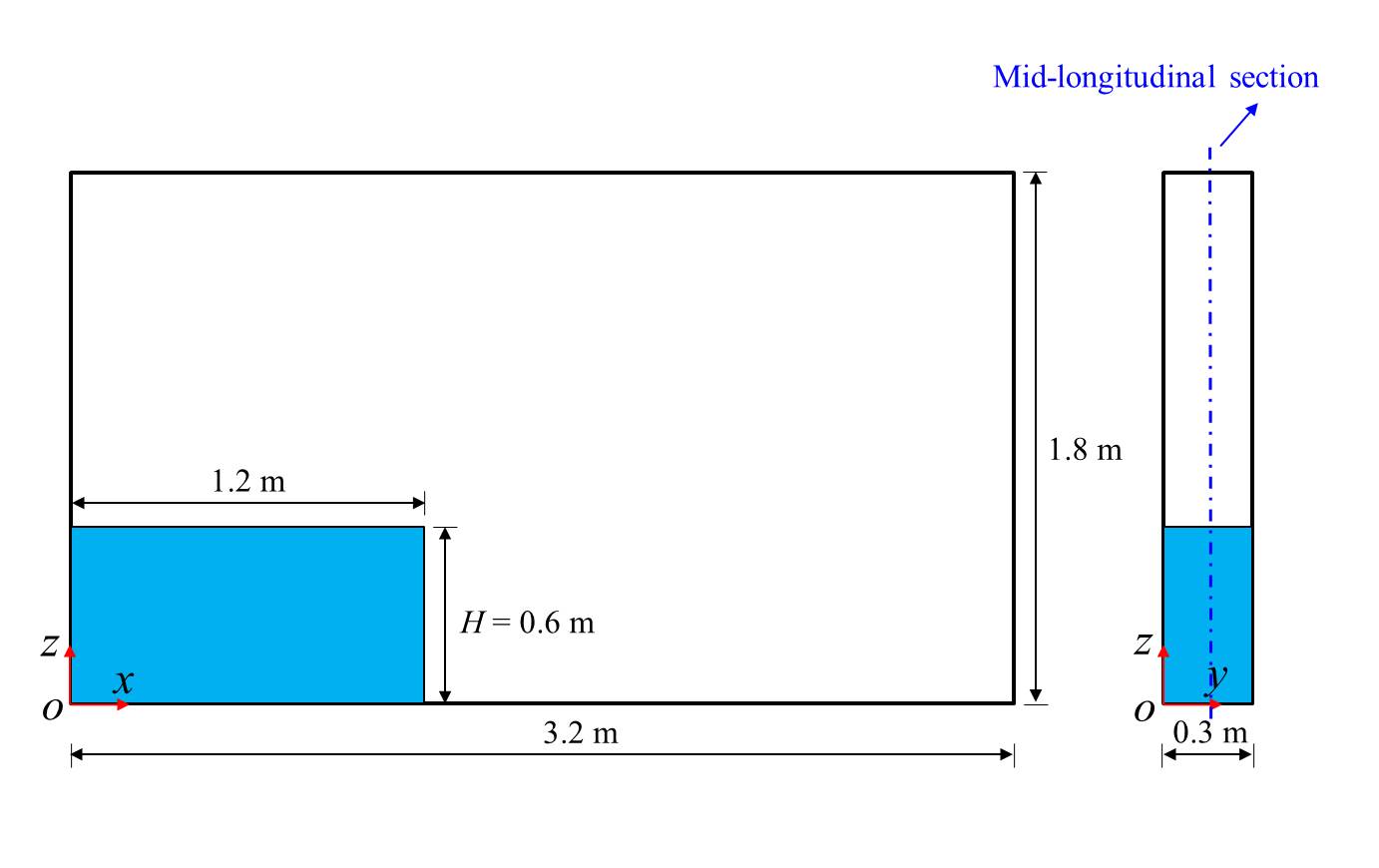}
\caption{Configuration of the dam-breaking. Left: $ x-z $ direction view. Right: $ y-z $ direction view.}
\label{fig:sketch-dambreaking}
\end{figure}

In this section, the present method is applied to the dam-breaking, and the detection results and detection time are compared with those of the Haque method \cite{HaqueDilts2007}, the Marrone method \cite{MarroneColagrossi2010}, and the Wang method \cite{WangMeng2019}. Because the detection capability for concave surfaces is improved and the detection accuracy is independent of the estimation of the normal vector of the particles, the detection accuracy of the present method is improved over the Haque method, the Marrone method, and the Wang method, and the free-surface particles are well detected using the present method. The free-surface particles detected by the present method are used as the reference free-surface particles in the subsequent comparison study. Figure~\ref{fig:sketch-dambreaking} shows the configuration of the 3D dam-breaking simulation. The water tank is 3.2 m long, 0.3 m wide, and 1.8 m high, and the water column is 1.2 m long, 0.3 m wide, and 0.6 m high. The initial inter-particle distance is 0.01 m. Ghost particle technology \cite{ColagrossiLandrini2003} is applied to impose boundary conditions. Furthermore, we record the total time of the first ten time steps and obtain the mean time of each method. All free-surface particle detections are performed on a computer with an i5-3470 processor at 3.20 GHz.

\begin{figure}[htbp]
\centering
\includegraphics[width=16cm]{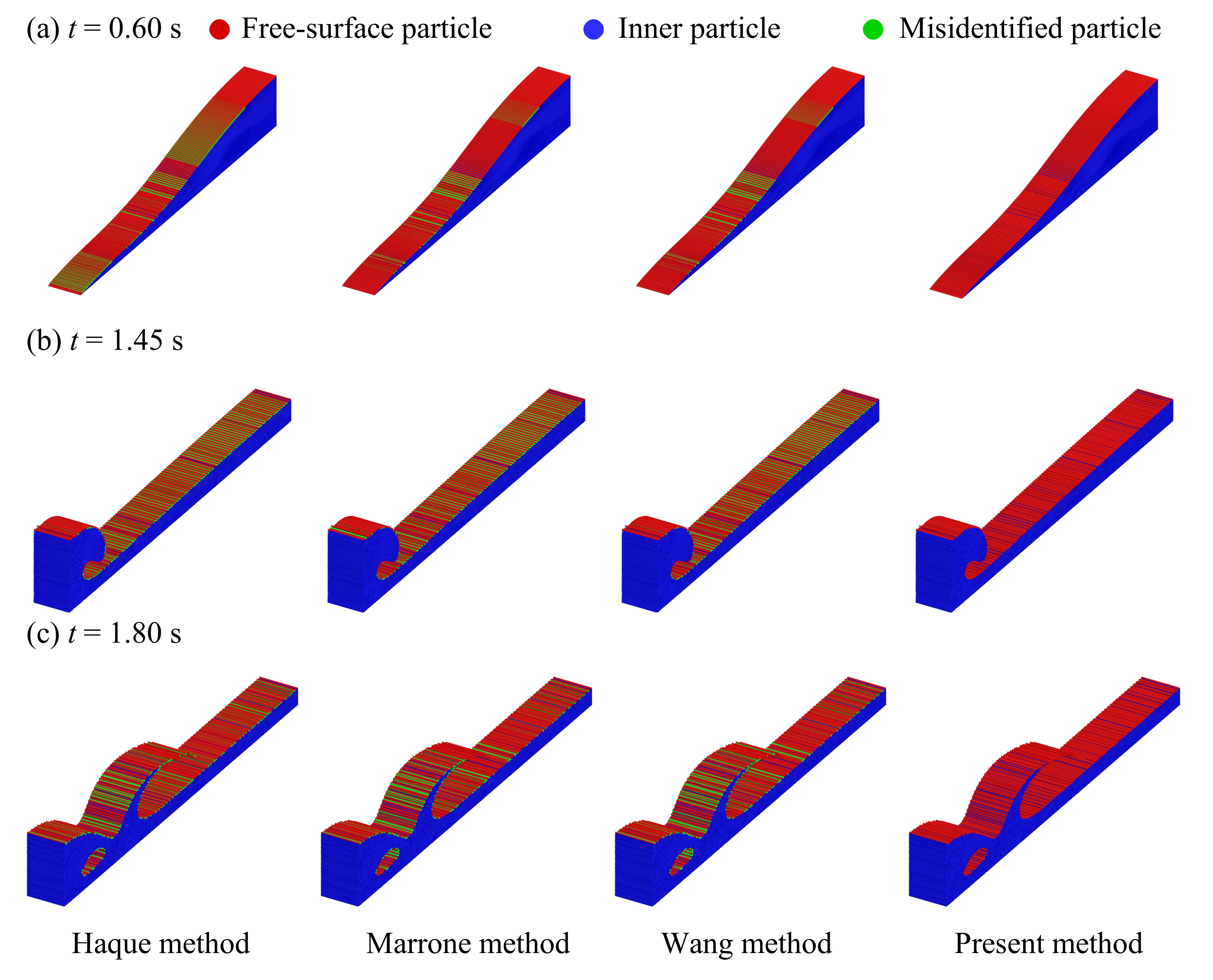}
\caption{Three-dimensional dam-breaking problem. The free-surface particles are detected by the Haque method \cite{HaqueDilts2007}, the Marrone method \cite{MarroneColagrossi2010}, the Wang method \cite{WangMeng2019}, and the present method. Red, blue, and green particles denote free-surface particles, inner particles, and misidentified particles, respectively.}
\label{fig:detection-result-dambreaking}
\end{figure}

\begin{table}[htbp]
\centering
\caption{Quantitative comparisons of the detection results of the three-dimensional dam-breaking problem. The first number denotes the number $ TP $ of reference free-surface particles, the second number denotes the number $ EP $ of misidentified particles, and the third number denotes the accuracy $ M $ of the method.}
\begin{tabular}
{m{12mm}<{\centering} m{33mm}<{\centering} m{33mm}<{\centering} m{33mm}<{\centering} m{31mm}<{\centering}}
\toprule
\tabincell{c}{} & \tabincell{c}{Haque method} & \tabincell{c}{Marrone method} & \tabincell{c}{Wang method} & \tabincell{c}{Present method}
\\ \midrule
\tabincell{c}{0.600 s} & 10440/3120/0.701 & 10440/1650/0.842 & 10440/1650/0.842 & 10440/0/1.000
\\ \midrule
\tabincell{c}{1.450 s} & 13380/4510/0.663 & 13380/3780/0.717 & 13380/3690/0.724 & 13380/0/1.000
\\ \midrule
\tabincell{c}{1.800 s} & 19380/5400/0.721 & 19380/4260/0.780 & 19380/4530/0.766 & 19380/0/1.000
\\ \bottomrule
\end{tabular}
\label{tab:dambreaking-3D}
\end{table}

\begin{figure}[htbp]
\centering
\includegraphics[width=16cm]{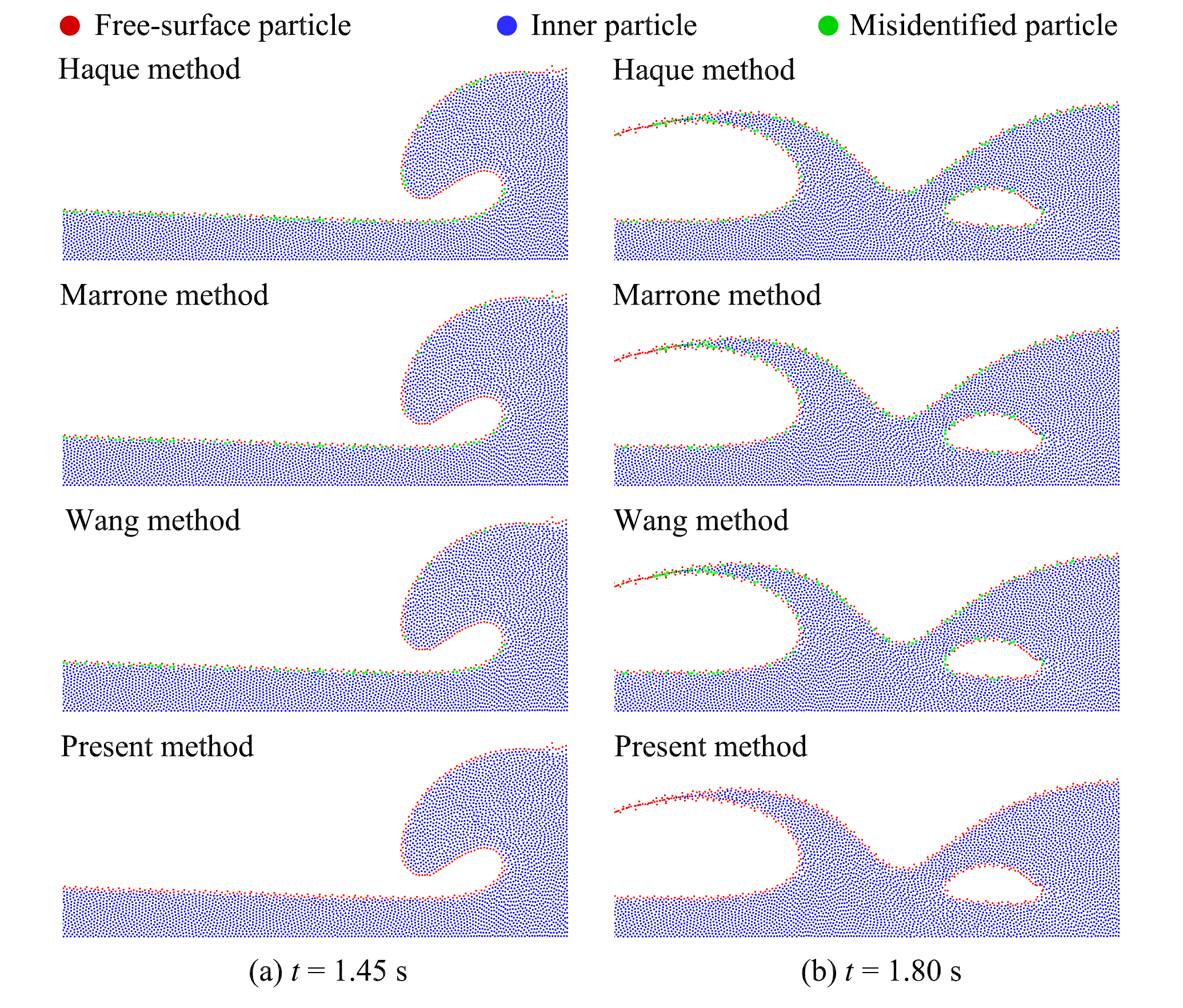}
\caption{Particle distributions of single particles in the right-hand fixed area of the mid-longitudinal section at 1.45 s and 1.80 s, and the free-surface particles are detected by the Haque method \cite{HaqueDilts2007}, the Marrone method \cite{MarroneColagrossi2010}, the Wang method \cite{WangMeng2019}, and the present method. Red, blue, and green particles denote free-surface particles, inner particles, and misidentified particles, respectively. Because the detection accuracy of the present method is improved over the Haque method, the Marrone method, and the Wang method, the free-surface particles detected by the present method are used as the reference free-surface particles.}
\label{fig:dambreaking-singlelayer}
\end{figure}

\begin{figure}[htbp]
\centering
\includegraphics[width=16cm]{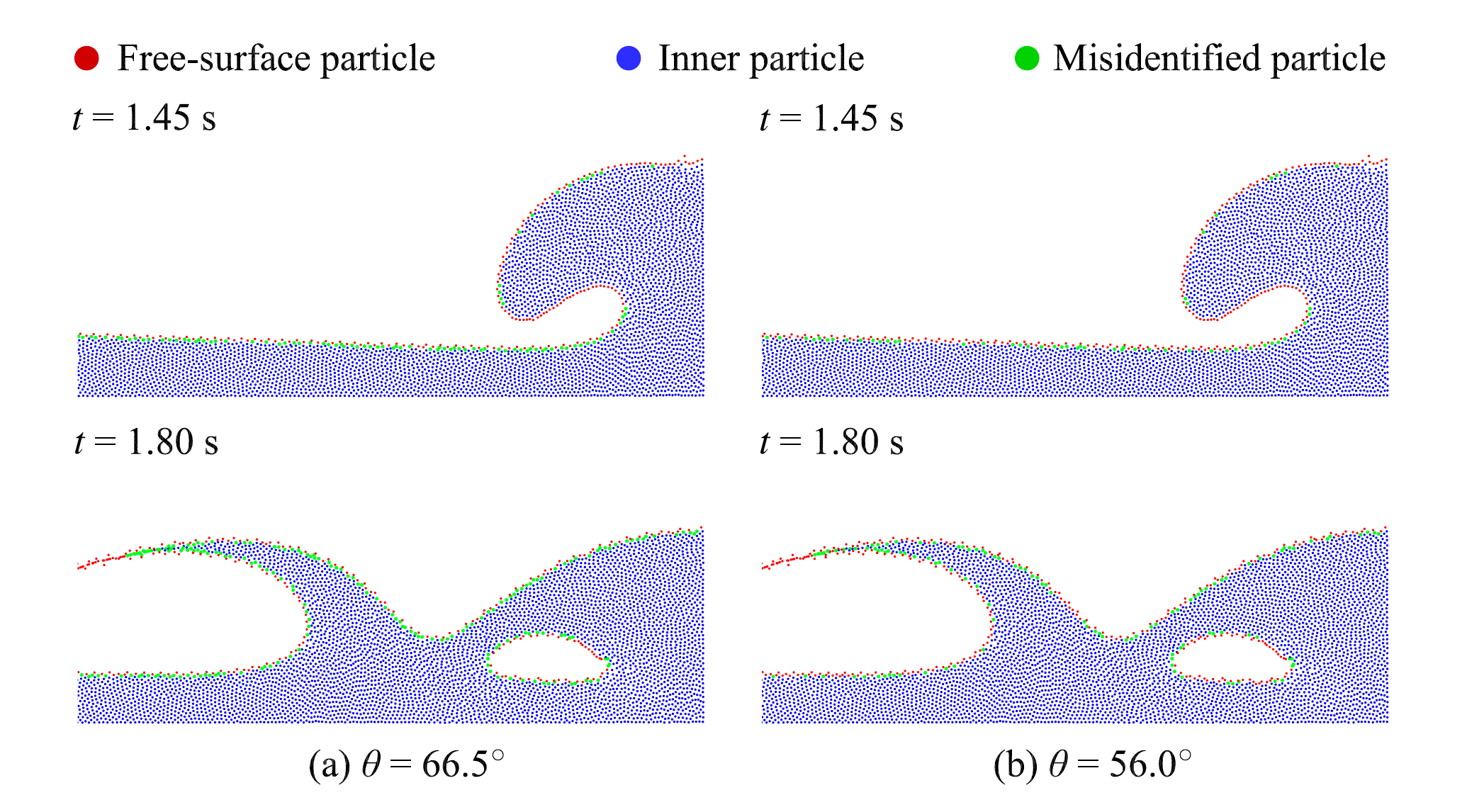}
\caption{Effect of the rotation angle of the cone on the detection results in the present method. Red, blue, and green particles denote free-surface particles, inner particles, and misidentified particles, respectively. The free-surface particles detected by the present method with a rotation angle of $ 45^{\circ} $ are used as the reference free-surface particles.}
\label{fig:result-dection-angle}
\end{figure}

Figure~\ref{fig:detection-result-dambreaking} shows the free-surface particle detection using the present method in dam-breaking, and the detection results are compared with those of the Haque method, the Marrone method, and the Wang method. Red, blue, and green particles denote free-surface particles, inner particles, and misidentified particles, respectively. The results show that the free-surface particles are well detected using the present method even when the particles are highly disordered in distribution. Several free-surface particles are misidentified as the inner particles owing to the reduced accuracy of the Haque method in detecting concave surfaces. Because the free-surface particles must be accurately detected by scanning neighboring particles along the normal direction in the second step of the Marrone method and the Wang method, several free-surface particles are misidentified as the inner particles owing to the inaccurate estimation of the normal vector in complex flows. Furthermore, quantitative comparisons of the detection results of the different methods are performed, as shown in Tables~\ref{tab:dambreaking-3D}. We measure the accuracy $ M $ of each method using the following expression:
\begin{equation}
M
=
\frac{TP-EP}{TP}
\end{equation}
where $ TP $ denotes the number of reference free-surface particles, and $ EP $ denotes the number of misidentified particles. As time goes by, the free-surface flows are gradually in a state of large deformation, and the detection accuracy of the Haque method, the Marrone method, and the Wang method all occur a significant decrease to just over seventy percent. Figure~\ref{fig:dambreaking-singlelayer} shows the particle distributions of single particles in the right-hand fixed area of the mid-longitudinal section at 1.45 s and 1.80 s, and the free-surface particles are detected by different methods. As illustrated in Fig.~\ref{fig:dambreaking-singlelayer}, the accuracy of the Haque method is reduced for detecting concave surfaces, thus, several free-surface particles on concave surfaces are misidentified as the inner particles. Because the free-surface particles must be accurately detected by scanning neighboring particles along the normal direction in the second step of the Marrone method and the Wang method, several free-surface particles have at least one neighboring particle in the scan region and are misidentified as the inner particles owing to the inaccurate estimation of the normal vector in complex flows. Although the Marrone method and the Wang method have the same procedure in the second step, i.e., the free-surface particles are further detected by a scan region along the normal direction of the candidate particle, the difference in the way the free-surface candidate particles are identified in the first step leads to a difference in the detection results in regions where only a small number of particles are present. Because the detection capability for concave surfaces influences the detection accuracy of free-surface particles, the effect of the rotation angle of the cone on the detection results in the present method is analyzed, as shown in Fig.~\ref{fig:result-dection-angle}. When the angle of the cone is $ 66.5^{\circ} $, which is half of the $ \theta_{\min} $ in Eq.~\refeq{eq:minimum-angle}, the detection results of free-surface particles are essentially the same as those of the Haque method. Owing to the limited detection capability for concave surfaces, only a small number of free-surface particles are detected. When the rotation angle is reduced to $ 56^{\circ} $, the number of misidentified free-surface particles decreases due to the higher detection capability for concave surfaces. When the angle of the cone is $ 45^{\circ} $, the free-surface particles are well detected using the present method. It is critical to note that the rotation angle of the cone cannot be too small, otherwise, several inner particles may be misidentified as the free-surface particles.

\begin{figure}[htbp]
\centering
\includegraphics[width=10cm]{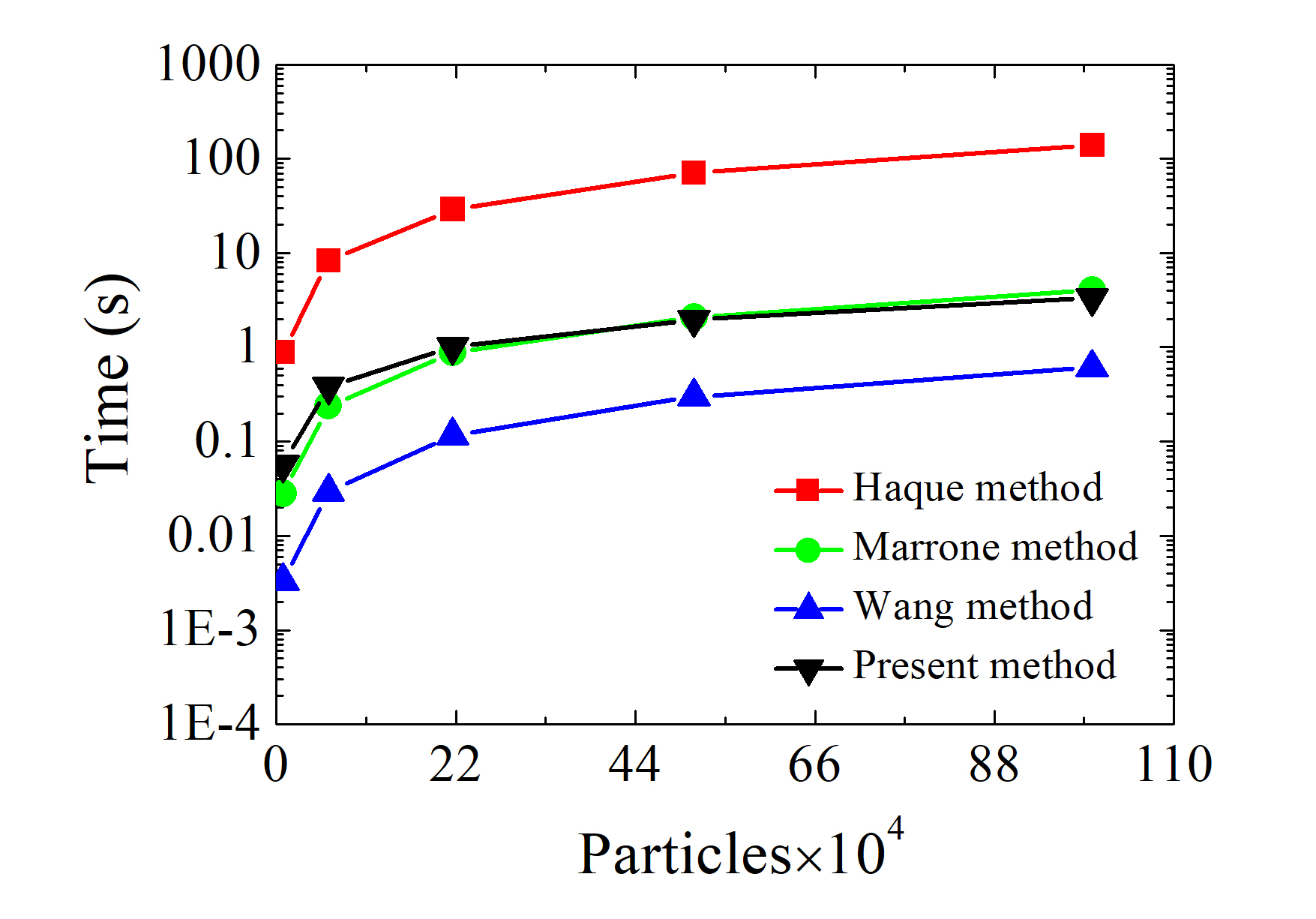}
\caption{The detection time of the Haque method \cite{HaqueDilts2007}, the Marrone method \cite{MarroneColagrossi2010}, the Wang method \cite{WangMeng2019}, and the present method with the number of particles.}
\label{fig:time-dection}
\end{figure}

Figure~\ref{fig:time-dection} gives the detection times of the Haque method, the Marrone method, the Wang method, and the present method. In the Haque method, because the circles obtained by the intersection of sphere $ S_i $ and its neighboring spheres $ S_j $ are assigned an uncertain normal direction, certain computations must be performed even if the circles do not intersect. When the particles are uniformly distributed, the upper limit of the computational time of the Haque method is $ O(MN^2) $, where $ M $ is the particle number and $ N $ is the average number of neighboring particles per particle. Therefore, it leads to huge computational costs. In the Marrone method, the minimum eigenvalue of the matrix $ \mathbf{L}^{-1} $ is computed for all particles in the first step, which takes most of the detection time. Meanwhile, the scanning by the ``umbrella-shaped'' zone along the normal direction is performed for a few particles in the second step, which takes only a small portion of the detection time. In the Wang method, the position divergence of the particle is computed for all particles in the first step, which takes only a small amount of time. Meanwhile, the scanning by the ``umbrella-shaped'' zone along the normal direction is performed for a few particles in the second step, which also takes a small amount of time. In the present method, the position divergence of the particle is computed for all particles in the first step, which takes only a small portion of the detection time. Meanwhile, continuous global scanning within the sphere is performed through the cone region for a few particles in the second step, which takes most of the detection time. Furthermore, because the circles given by the intersection of a sphere $ S_{i} $ and its neighboring cones $ C_{j} $ have the same radius and outward normal direction in the continuous global scanning, only when the angle $ \varphi_{ijk} $ is greater than $ 0^{\circ}$ and less than $ 2\theta $, circles $ C_{ij} $ and $ C_{ik} $ intersect at two points, and the point $ \boldsymbol{x}_{ijk} $ of the intersection of the planes of the circles $ C_{ij} $, $ C_{ik} $ and $ C_{ijk} $ is unique and can be calculated directly. When the particles are uniformly distributed, because the angle $ \theta $ is set to $ 45^{\circ} $, only half of the neighboring circles $ C_{ik} $ intersect the circle $ C_{ij} $, and the upper limit of the computational time of the present geometrical method is $ O(0.5MN^2) $. Therefore, the detection time of a single particle in the second step of the present method is significantly reduced compared with the Haque method. When the particle number is small, the detection time of the Marrone method is shorter than that of the present method. When the particle number increases, the detection time of the Marrone method increases approximately linearly with respect to the number of particles. However, in the present method, the proportion of particles detected in the second step compared to the total number of particles gradually decreases with the number of particles, hence, the increasing rate of the detection time for the present method gradually decreases. When the particle number is one million, the detection time is 139.79 s for the Haque method, 4.04 s for the Marrone method, 0.65 s for the Wang method, and 3.34 s for the present method. Compared with the Haque method and the Marrone method, the detection time of the present method is reduced by 97.61\% and 17.33\%, respectively. Compared with the Wang method, the detection time of the present method is increased by 413.85\%.

\begin{figure}[htbp]
\centering
\includegraphics[width=16cm]{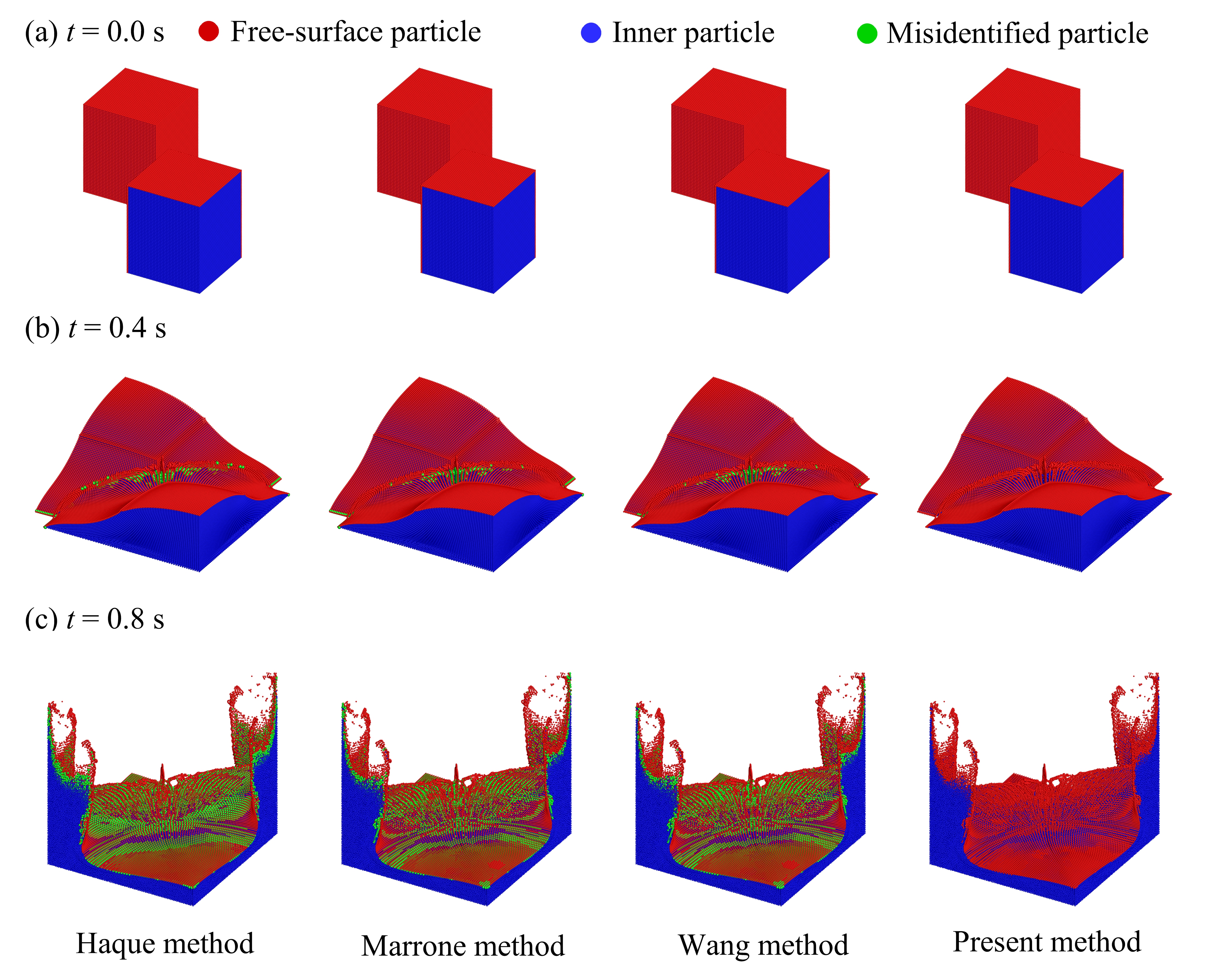}
\caption{Three-dimensional dam-breaking problem of two liquid columns at opposite corners of a container with a square base of size 2.4 and height 3.5. The free-surface particles are detected by the Haque method \cite{HaqueDilts2007}, the Marrone method \cite{MarroneColagrossi2010}, the Wang method \cite{WangMeng2019}, and the present method. Red, blue, and green particles denote free-surface particles, inner particles, and misidentified particles, respectively.}
\label{fig:dambreaking-twoliquid}
\end{figure}

\begin{figure}[htbp]
\centering
\includegraphics[width=16cm]{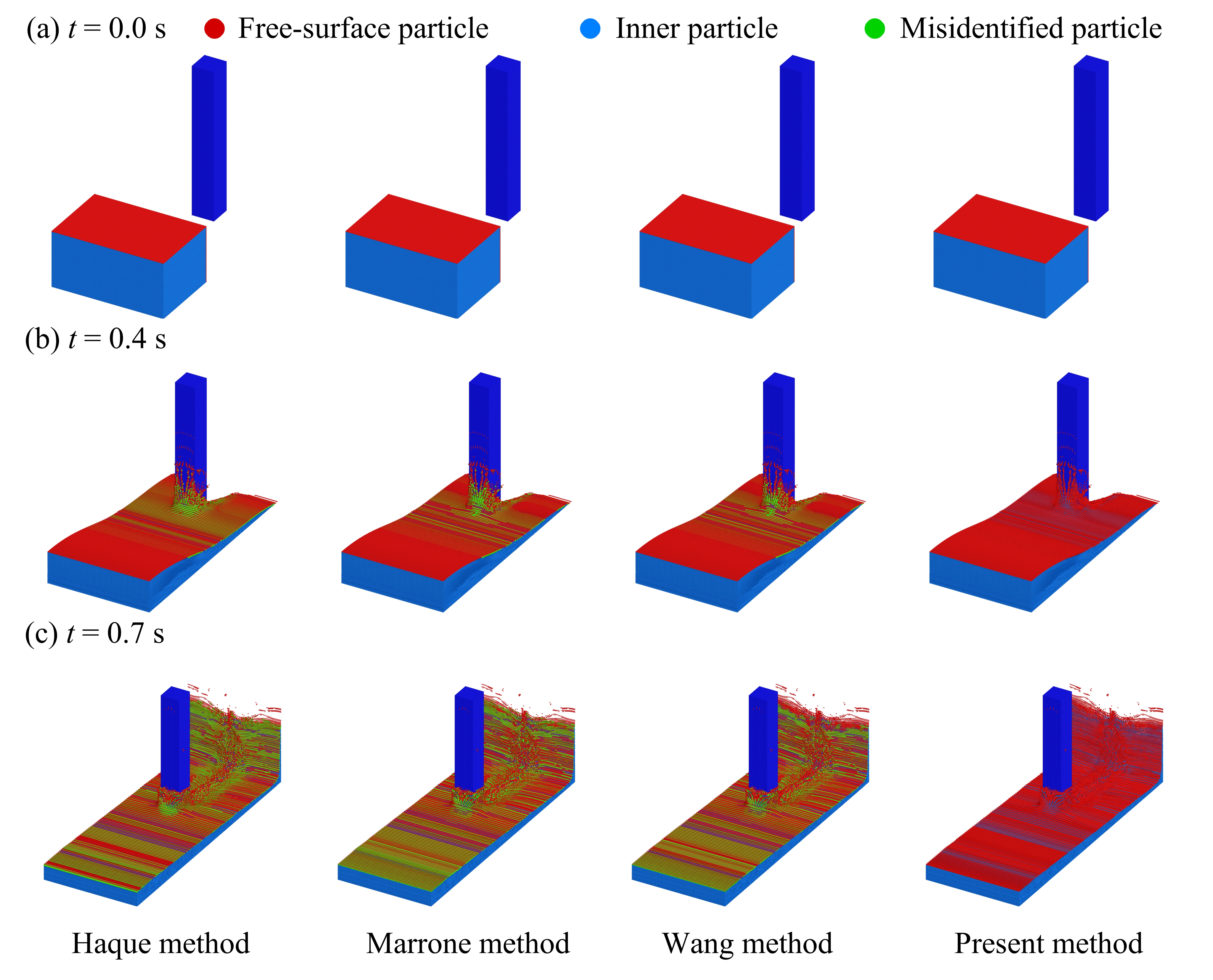}
\caption{Impact against a rigid obstacle after a dam-break. The free-surface particles are detected by the Haque method \cite{HaqueDilts2007}, the Marrone method \cite{MarroneColagrossi2010}, the Wang method \cite{WangMeng2019}, and the present method. Red, blue, and green particles denote free-surface particles, inner particles, and misidentified particles, respectively.}
\label{fig:dambreaking-obstacle}
\end{figure}

Furthermore, Figures~\ref{fig:dambreaking-twoliquid} and ~\ref{fig:dambreaking-obstacle} depict the free-surface particle detection using the present method in complex flows resulting from the impact of a double dam-breaking and the impact of a single dam-breaking against a rigid tall obstacle, respectively. In each case, the detection results are compared with those of the Haque method, the Marrone method, and the Wang method. The results show that the free-surface particles are well detected using the present method even when the particles are highly disordered in distribution. Quantitative comparisons of the detection results of the different methods are also performed, as shown in Tables~\ref{tab:dambreaking-twoliquid} and~\ref{tab:dambreaking-obstacle}. At the initial moment, all free-surface particles are accurately detected by the different methods because of the uniform distribution of particles. However, as time goes by, the free-surface flows are gradually in a state of large deformation, and the detection accuracy of the Haque method, the Marrone method, and the Wang method all occur a significant decrease to just over sixty percent.

\begin{table}[htbp]
\centering
\caption{Quantitative comparisons of the detection results of the three-dimensional dam-breaking problem of two liquid columns at opposite corners of a container. The first number denotes the number $ TP $ of reference free-surface particles, the second number denotes the number $ EP $ of misidentified particles, and the third number denotes the accuracy $ M $ of the method.}
\begin{tabular}
{m{12mm}<{\centering} m{33mm}<{\centering} m{33mm}<{\centering} m{33mm}<{\centering} m{31mm}<{\centering}}
\toprule
\tabincell{c}{} & \tabincell{c}{Haque method} & \tabincell{c}{Marrone method} & \tabincell{c}{Wang method} & \tabincell{c}{Present method}
\\ \midrule
\tabincell{c}{0.000 s} & 16882/0/1.000 & 16882/0/1.000 & 16882/0/1.000 & 16882/0/1.000
\\ \midrule
\tabincell{c}{0.400 s} & 17274/904/0.948 & 17274/712/0.959 & 17274/657/0.962 & 17274/0/1.000
\\ \midrule
\tabincell{c}{0.800 s} & 38281/14922/0.610 & 38281/12025/0.686 & 38281/12225/0.681 & 38281/0/1.000
\\ \bottomrule
\end{tabular}
\label{tab:dambreaking-twoliquid}
\end{table}

\begin{table}[htbp]
\centering
\caption{Quantitative comparisons of the detection results of the impact against a rigid obstacle after a dam-break. The first number denotes the number $ TP $ of reference free-surface particles, the second number denotes the number $ EP $ of misidentified particles, and the third number denotes the accuracy $ M $ of the method.}
\begin{tabular}
{m{12mm}<{\centering} m{33mm}<{\centering} m{33mm}<{\centering} m{33mm}<{\centering} m{31mm}<{\centering}}
\toprule
\tabincell{c}{} & \tabincell{c}{Haque method} & \tabincell{c}{Marrone method} & \tabincell{c}{Wang method} & \tabincell{c}{Present method}
\\ \midrule
\tabincell{c}{0.000 s} & 16680/0/1.000 & 16680/0/1.000 & 16680/0/1.000 & 16680/0/1.000
\\ \midrule
\tabincell{c}{0.400 s} & 37647/12207/0.676 & 37647/10057/0.733 & 37647/9869/0.738 & 37647/0/1.000
\\ \midrule
\tabincell{c}{0.700 s} & 71173/24405/0.657 & 71173/26106/0.633 & 71173/23495/0.670 & 71173/0/1.000
\\ \bottomrule
\end{tabular}
\label{tab:dambreaking-obstacle}
\end{table}

\subsection{Visualization tool}

\begin{figure}[htbp]
\centering
\includegraphics[width=16cm]{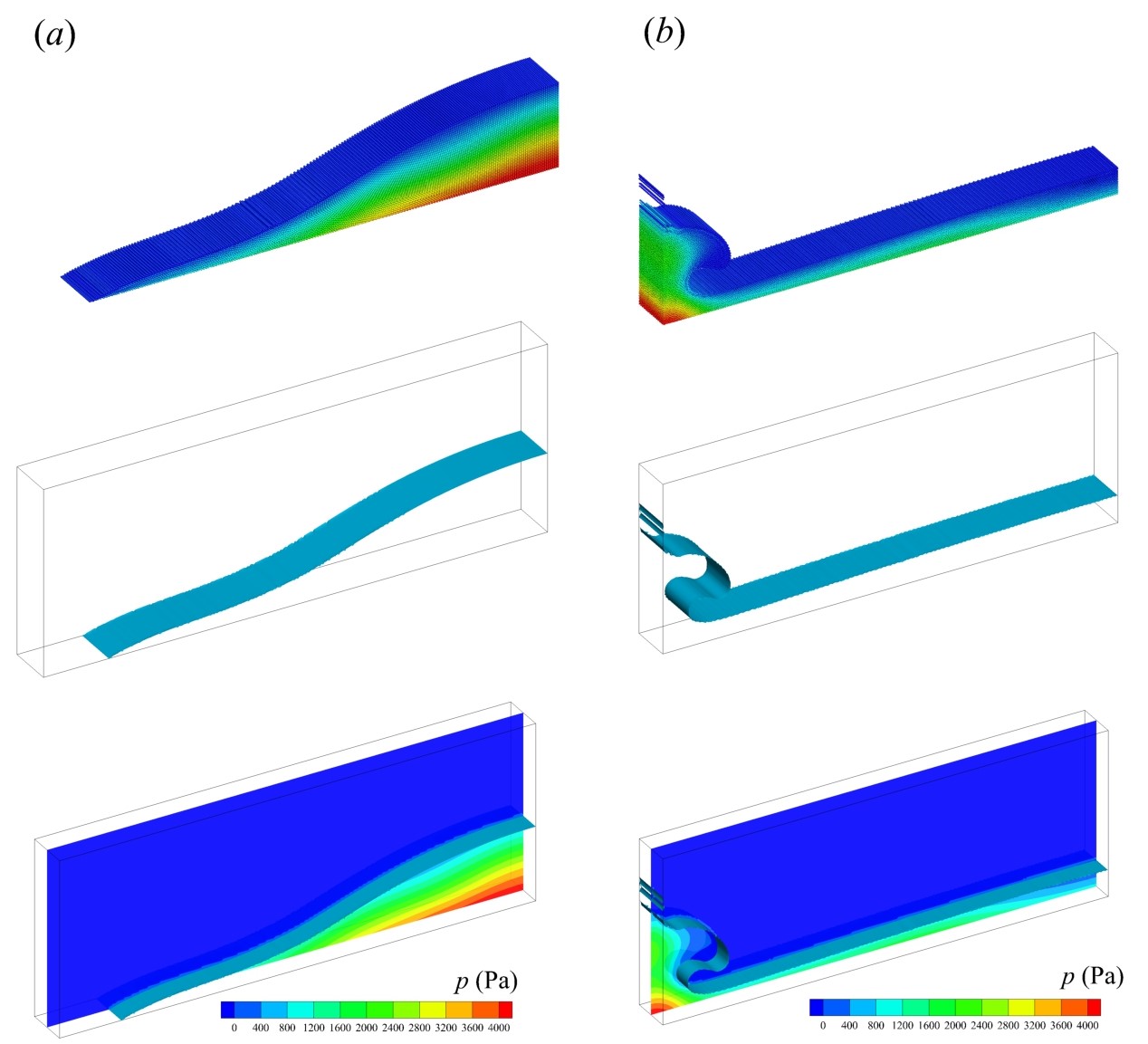}
\caption{Three-dimensional dam-breaking. (a) 0.52 s; (b) 1.42 s. Top: particle distributions. Middle: free surface. Bottom: pressure contours interpolated on the mid-longitudinal section.}
\label{fig:visualization-dambreaking}
\end{figure}

The SPH method always outputs computation results in the form of discrete points, making it difficult to obtain slices, iso-surfaces, contour maps, and streamlines. The aforementioned analysis, however, can be conducted directly using existing tools if the particle data can be interpolated from the neighboring regular meshes. Marrone et al. \cite{MarroneColagrossi2010} developed the following method for analyzing free-surface flows in detail.

(1) Detect free-surface particles.

(2) Construct regular meshes with spatial resolution $ dx $ containing all computational domains. For each mesh $ M $, search for the nearest free-surface particle $ F_M $, and obtain $ d_{MF_M} $ as:
\begin{equation}
    d_{MF_M} = \left( \boldsymbol{x}_{F_M} - \boldsymbol{x}_M \right) \cdot \boldsymbol{n}_{F_M},
\end{equation}
where $ \boldsymbol{n}_{F_M} $ is the normal vector of particle $ F_M $. Then, a level-set function $ \phi \left( \boldsymbol{x}_M \right) $ on each mesh can be defined as follows:
\begin{equation}
    \phi \left( \boldsymbol{x}_M \right)
    =
    \left\{
    \begin{array}{ll}
    -1,
    &
    d_{MF_M}\leqslant -2h \medskip \\
    d_{MF_M}/2h,
    &
    -2h<d_{MF_M}<2h \medskip \\
    1,
    &
    d_{MF_M}\geqslant 2h
    \end{array}
    \right.
\end{equation}
The level-set function is negative outside the fluid, zero on the free surface where $ \phi = -dx/4h $, and positive inside the fluid.

(3) For each mesh $ M $, the physical quantity on the mesh can be interpolated from the neighboring scattered point data using the moving least squares (MLS) method \cite{FriesMatthies2004}.

Figure~\ref{fig:visualization-dambreaking} depicts the 3D dam-breaking at 0.52 s and 1.42 s, and the free-surface particles are detected by the present method. The top shows particle distributions, the middle shows the free surface, and the bottom shows pressure contours on the mid-longitudinal section. It can be seen that the visualization tools can clearly describe the surface position and internal flow.

\subsection{Particle shifting technology}

\begin{figure}[htbp]
\centering
\includegraphics[width=16cm]{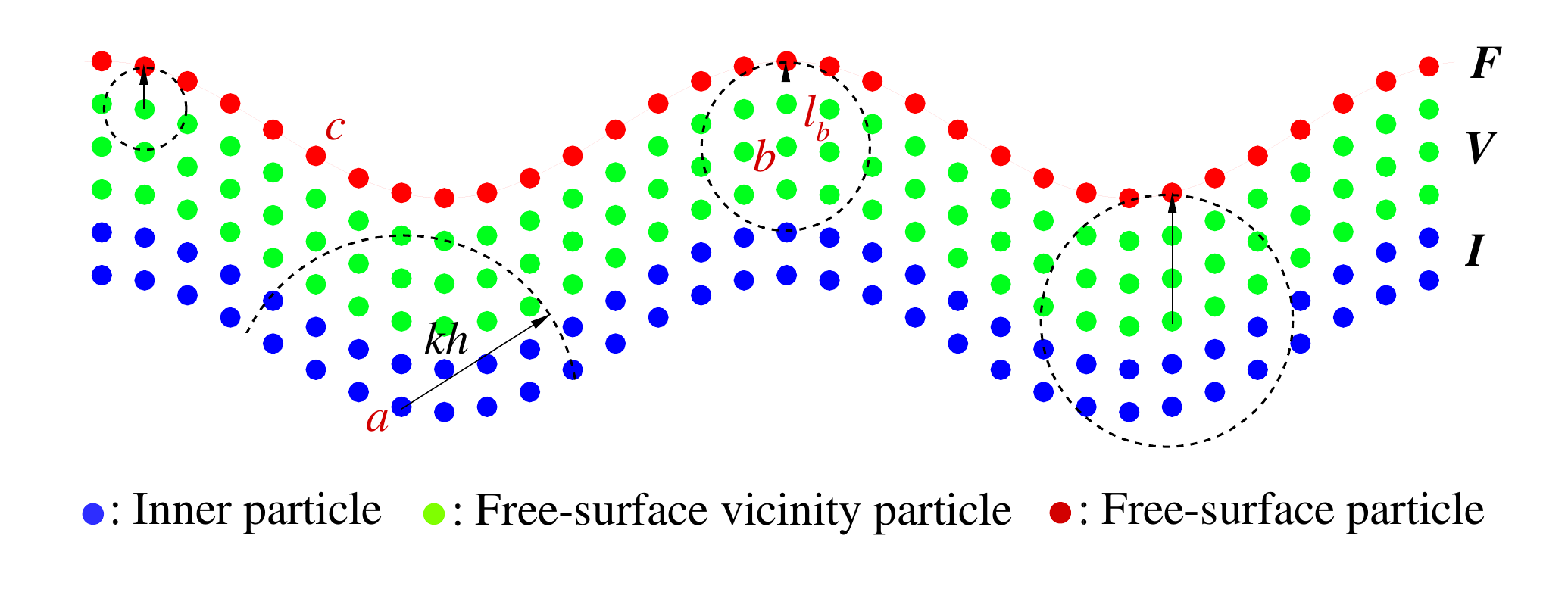}
\caption{Sketch of the PST. For the inner particle $ a $, all neighboring particles participate in the calculation of the shifting vector. For the free-surface vicinity particle $ b $, only the neighboring particles whose distance from the particle is less than $ l_b $ participate in the calculation of the shifting vector. For the free-surface particle $ c $, the shifting distance is zero.}
\label{fig:sketch-pst}
\end{figure}

PST has become a popular technology in recent years for improving the robustness and accuracy of the SPH method \cite{XuStansby2009, LindXu2012, SunColagrossi2017, WangMeng2019, LyuSun2022}. Xu et al. \cite{XuStansby2009} first developed the PST. In this method, the particles are shifted slightly across streamlines to avoid the extreme bunching and stretching of particles. When the PST is applied to free-surface flows, special treatment of the free-surface particles is required to maintain the kinematic boundary condition \cite{SunColagrossi2017, HuangLong2019, WangMeng2019, LyuSun2022}. Wang et al. \cite{WangMeng2019} proposed an improved PST that enables accurate shifting of particles near the free surface. This method is depicted in Fig. 17 and is implemented as follows:

(1) Detect free-surface particles.

(2) For free-surface vicinity particles, record the distance $ l $ between them and the nearest free-surface particle.

(3) For inner particles, all neighboring particles participate in the calculation of the shifting vector. For free-surface vicinity particles, only neighboring particles whose distance from the particle is less than $ l $ participate in the calculation of the shifting vector. For free-surface particles, the shifting distance is zero.

\begin{figure}[htbp]
\centering
\includegraphics[width=16cm]{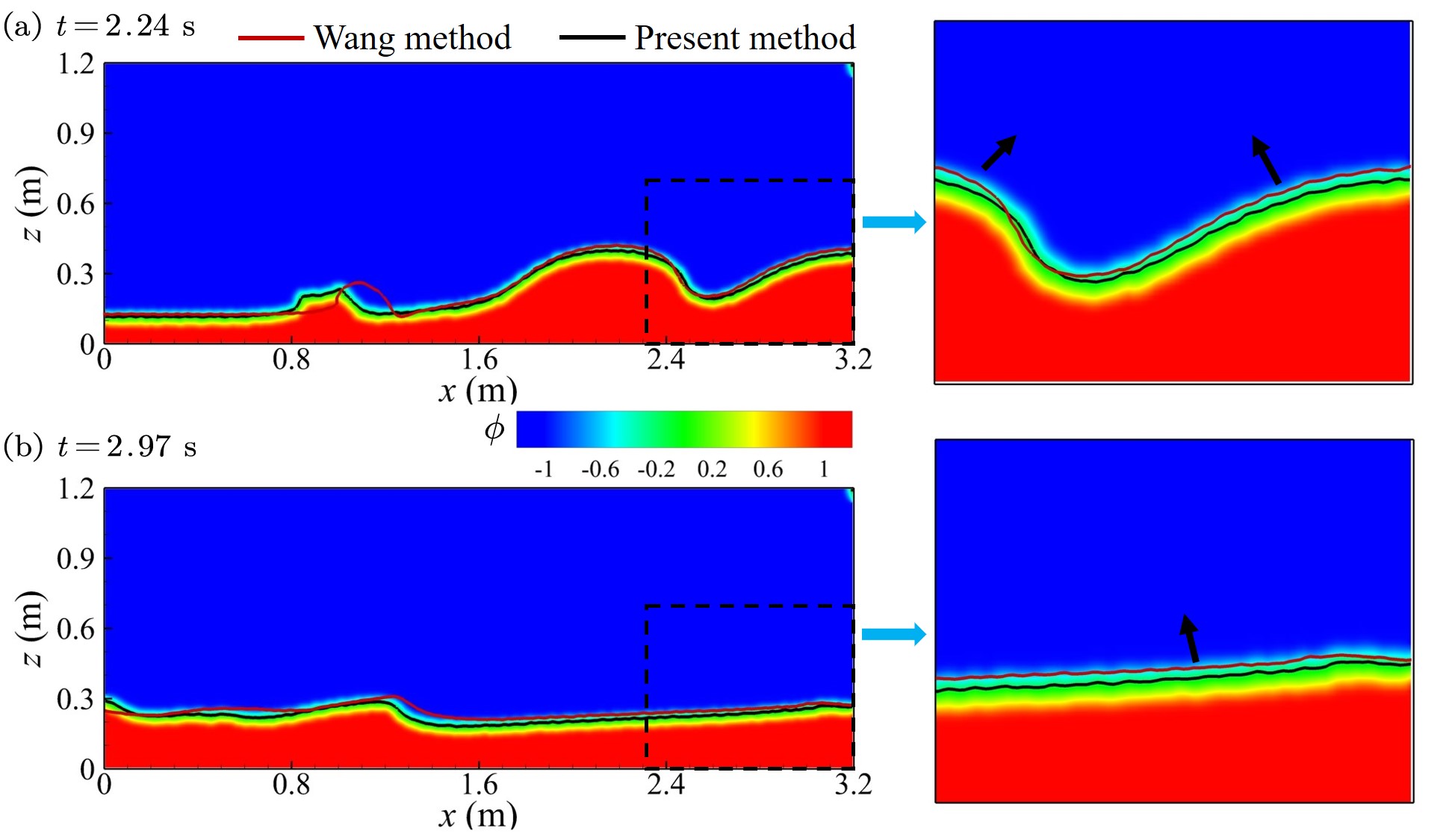}
\caption{Visualization results of the single-layer particles on the mid-longitudinal sector at 2.24 s and 2.97 s, and the free-surface particles in PST are detected by the Wang method \cite{WangMeng2019} and the present method, respectively. The red line denotes the free surface obtained by applying the Wang method to PST, and the black line denotes the free surface obtained by applying the present method to PST. The arrow indicates the direction of free surface diffusion.}
\label{fig:PST-displacement}
\end{figure}

Figure~\ref{fig:PST-displacement} shows the visualization results of the single-layer particles on the mid-longitudinal sector at 2.24 s and 2.97 s, and the free-surface particles in PST are detected by the Wang method and the present method, respectively. The results show that the diffusion of the free surface obtained by applying the Wang method to PST is observed compared with the results obtained by the present method. Because the free-surface particles must be accurately detected by scanning neighboring particles along the normal direction in the second step of the Wang method, several free-surface particles are misidentified as the inner particles owing to the inaccurate estimation of the normal vector in complex flows. This part of misidentified particles obtains a PST displacement towards the free surface, therefore, the number of particles in the free-surface region gradually increases and the free surface diffusion occurs. Because the detection accuracy is independent of the estimation of the normal vector of particles, the free-surface particles are accurately detected by the present method, and the PST displacement of the free-surface particles is zero, therefore, there is no particle aggregation in the free-surface region and no diffusion on the free surface.

%===============================================================================
\section{Conclusions}\label{sec:conclusion}

A high-accuracy three-dimensional surface detection algorithm is studied for the SPH method. First, a geometrical method for free-surface particle detection is developed to improve detection accuracy in complex flows. This method detects free-surface particles by performing continuous global scanning within the sphere of a particle through a cone region. The vertex of the cone corresponds to the particle position. If there exists a cone region with no neighboring particles, the particle is identified as a free-surface particle. Then, based on the proposed geometrical method, a two-step semi-geometrical method is developed to reduce computational costs. In the first step, the particles near the free surface are found by position divergence. In the second step, the free-surface particles are accurately detected using the proposed geometrical method.

The accuracy and robustness of the proposed method are demonstrated via four tests. The test demonstrating the surface detection of the free surface with periodic perturbations shows that the detection accuracy of the proposed method is significantly improved compared with the traditional methods because the detection capability for concave surfaces is enhanced and the detection accuracy is independent of the estimation of the normal vector of particles. The surface detection of 3D dam-breaking tests shows that the proposed method can well detect the free-surface particles in complex flows but with no increase in computational costs compared with the traditional methods. The results of the tests on visualization tools and the particle shifting technology show that the proposed method can be applied to various fields with satisfactory results and has a wide range of application prospects. In conclusion, all tests demonstrate that the proposed method is effective and robust for complex flows.

\section*{Acknowledgments}

This work is supported by the National Natural Science Foundation of China under Grant Nos.
U1530261, 11772065, and the Science Challenge Project, China (Grant No. TZ2016001).

\section*{Reference}

  \bibliographystyle{elsarticle-num}
  \bibliography{myBib_SPH}




%%\bibitem{}

%%\end{thebibliography}

\end{document}